\documentclass[12pt]{article}

\date{ }

\usepackage{amscd,amssymb,amsmath,latexsym,enumerate}
\usepackage[mathscr]{euscript}
\usepackage{epsfig}
\usepackage{fancybox}
\usepackage{verbatim}

\usepackage{tikz}
\usepackage{tikz-cd}
\usepackage{xcolor}
\definecolor{MyBlue}{cmyk}{1,0.13,0,0.63}
\definecolor{MyGreen}{cmyk}{0.91,0,0.88,0.52}
\newcommand{\mylinkcolor}{MyBlue}
\newcommand{\mycitecolor}{MyGreen}
\newcommand{\myurlcolor}{black}

\usepackage{hyperref}
\hypersetup{%
  bookmarksnumbered=true,bookmarksopen=false,%
  plainpages=false,
  linktocpage=true,%
  colorlinks=true,breaklinks=true,%
  linkcolor=\mylinkcolor,citecolor=\mycitecolor,urlcolor=\myurlcolor,%
  pdfpagelayout=OneColumn,%
  pageanchor=true,%
}

\usepackage{color}
\newcommand{\red}{\textcolor[rgb]{0.99,0.00,0.00}}

\title{Space versus energy oscillations of Pr\"ufer phases \\ for matrix Sturm-Liouville and Jacobi operators}

\author{Hermann Schulz-Baldes and Liam Urban
\\
\\
{\small Department Mathematik, Friedrich-Alexander-Universit\"at Erlangen-N\"urnberg, Germany}
}

\newtheorem{theo}{Theorem}

\newtheorem{proposi}{Proposition}

\newtheorem{coro}{Corollary}

\newcommand{\CM}{{\mathbb C}}

\newcommand{\RM}{{\mathbb R}}
\newcommand{\SM}{{\mathbb S}}

\newcommand{\LM}{{\mathbb L}}

\newcommand{\Ee}{{\mathcal E}}
\newcommand{\Pp}{{\mathcal P}}

\newcommand{\Dd}{{\mathcal D}}

\newcommand{\Gg}{{\mathcal G}}

\newcommand{\Vv}{{\mathcal V}}
\newcommand{\Ss}{{\mathcal S}}
\newcommand{\Oo}{{\mathcal O}}
\newcommand{\Tt}{{\mathcal T}}

\newcommand{\Nn}{{\mathcal N}}
\newcommand{\Mm}{{\mathcal M}}

\newcommand{\Jj}{{\mathcal J}}

\newcommand{\Hh}{{\mathcal H}}

\newcommand{\one}{{\bf 1}}

\newcommand{\Tr}{\mbox{\rm Tr}}

\newcommand{\SF}{{\rm Sf}}

\newcommand{\Ker}{{\rm Ker}} 
 
\newcommand{\Ran}{{\rm Ran}}

\newcommand{\diag}{{\rm diag}}

\newcommand{\Morse}{i}
\newcommand{\MorseC}{i_{\mbox{\rm\tiny C}}}

\begin{document}

\maketitle

\begin{abstract}
This note considers Sturm oscillation theory for regular matrix Sturm-Liouville operators on finite intervals and for matrix Jacobi operators. The number of space oscillations of the eigenvalues of the matrix Pr\"ufer phases at a given energy, defined by a suitable lift in the Jacobi case, is shown to be equal to the number of eigenvalues below that energy. This results from a positivity property of the Pr\"ufer phases, namely they cannot cross $-1$ in the negative direction, and is also shown to be closely linked to the positivity of the matrix Pr\"ufer phase in the energy variable. The theory is illustrated by numerical calculations for an explicit example.

\vspace{.1cm}

\hfill
\end{abstract}


\section{Introduction}

Classical Sturm oscillation theory states that the number of oscillations of the fundamental solutions of a regular Sturm-Liouville equation at energy $E$ and over a (possibly rescaled) interval $[0,1]$ is equal to the number of eigenvalues of the Sturm-Liouville operator on the interval with energy less than or equal to $E$. This is also given by the rotation of the Pr\"ufer phase $e^{\imath \theta^E_x}$ in the spatial coordinate $x$. Alternatively, it is also equal to the rotation of the Pr\"ufer phase $e^{\imath \theta^e_1}$ at the end point $1$ of the interval, when the energy is varied in $e\in(-\infty,E]$.  A nice historic account of these facts is given in \cite{AHP}. 

\vspace{.2cm}

Matrix Sturm-Liouville equations are of the same form as the classical ones, but the coefficient functions now take values in the square matrices of a given fixed size. They are not only of intrinsic mathematical interest, but also of great relevance for numerous applications, such as the Jacobi equation for closed geodesics (however, with periodic boundary conditions),  mathematical physics, and many more, resulting in an abundant mathematical literature dating back many decades. Morse developed a variational approach to the study of closed geodesics \cite{Mor}, that was further extended by Bott \cite{Bot}. Intersection theory of Lagrangian planes for the associated eigenvalue problem was developed by Lidskii \cite{Lid} and Bott \cite{Bot}, see also the follow-up by Bott's student Edwards \cite{Edw}. The matrix Pr\"ufer phase was used in these works, albeit not under this name. It is a unitary matrix $U^E(x)$ the definition of which is recalled below. When stemming from a matrix Sturm-Liouville equation, it depends on energy and position. Its eigenvalues (on the unit circle) are also called Pr\"ufer phases. Actually the matrix Pr\"ufer phase is merely a global chart for the (hermitian symplectic) Lagrangian planes as given by the fundamental solution of the Sturm-Liouville equation with a fixed left boundary condition (which is a Lagrangian plane). The matrix Pr\"ufer phase allows to read off the dimension of the intersection of the Lagrangian plane of the solution and the right boundary condition. This intersection theory is essentially due to Bott and was further developed and applied by Maslov \cite{Mas}. The associated intersection number should therefore be called the Bott-Maslov index. Its relevance for Sturm oscillation theory was stressed by Arnold \cite{Arn}, see also \cite{SB} where all the above is explained in detail.

\vspace{.2cm}

Positivity properties of the Pr\"ufer phases in its parameters $E$ and $x$ are crucial elements of oscillation theory.  Bott  proved in \cite{Bot}, by an argument that is essentially reproduced in Theorem~\ref{theo-osci2} below, that the Pr\"ufer phases (unit eigenvalues of the matrix Pr\"ufer phase) always rotate in the positive sense as a function of the energy $E$.  This type of positivity is very robust and holds also for more general Hamiltonian systems (not stemming from a Sturm-Liouville equation), for block Jacobi operators \cite{SB,SB2} and even in a setting with infinite dimensional fibers \cite{GSV}. On the other hand, Lidski argued \cite{Lid} that the Pr\"ufer phases rotate in the positive sense as functions of the position $x$ as well, provided that a certain positivity property of the matrix potential holds. This was later on refined by Atkinson \cite{Atk} for a particular class of Hamiltonian systems, and for more general ones by Coppel \cite{Cop}, see also the book by Reid \cite{Rei}.  For general potentials entering the Sturm-Liouville operator, this monotonicity of the Pr\"ufer phases in $x$ simply does not hold, even for scalar Sturm-Liouville operators. This is clearly visible in the numerical example in Section~\ref{sec-Numerics} below. One contribution of this note (going slightly beyond \cite{Cop,Rei}) is to show that for any matrix Sturm-Liouville operator there is nevertheless a positivity in $x$, albeit in the following restricted sense: the Pr\"ufer phases always pass through $-1$ in the positive sense (Theorem~\ref{theo-osci3}). This fact is of crucial importance for the eigenvalue counting and allows to reconcile space and energy oscillations of the Pr\"ufer phases. We also stress the geometric aspects of the problem and thus offer a modern perspective on the above classical results.

\vspace{.2cm}

Jacobi matrices are the discrete analogues of Sturm-Liouville operators. The eigenvalue calculation can be done via Pr\"ufer phases which have the same positivity properties in the energy, also in the matrix-valued case \cite{SB,SB2}. For positivity in space and the Sturm oscillation theory, considerable care is needed though as it depends on the choice of interpolation between the discrete points in space. Building on a detailed spectral analysis of the transfer matrix in the generalized Lorentz group, Section~\ref{sec-Jacobi} shows how to construct a Sturm-Liouville operator with piecewise continuous coefficients associated to the Jacobi matrix, and how the space oscillations of its Pr\"ufer phase are linked to the spectral properties of the Jacobi matrix. 

\vspace{.2cm}

The paper continues in Section~\ref{sec-MatrixSL} by recalling the definition of a matrix Sturm-Liouville equation and the selfadjoint operator given by separate boundary conditions at the ends of the interval (periodic boundary conditions are discussed in \cite{Bot,SB2} and are dealt with similarly). Section~\ref{sec-MatrixPruefer} introduces the matrix Pr\"ufer phase and states how it can be used for the eigenvalue calculation of the Sturm-Liouville operator. Section~\ref{sec-EnergyOsci} proves the positivity of the matrix Pr\"ufer phase in energy. In the following Sections~\ref{sec-SpaceOsci} and \ref{sec-Asym}, the space oscillations and asymptotics of the Pr\"ufer phase are analyzed. Section~\ref{sec-Numerics} contains a numerical example in order to illustrate the theoretical results. Section~\ref{sec-Jacobi} then shows how to transpose the framework and some of the results to matrix-valued Jacobi matrices. Finally Sections~\ref{sec-PrueferJacobi} and \ref{sec-PrueferJacobi2} provide two separate interpolations in the Pr\"ufer matrices in the space variable that allow to compare space and energy oscillations also for Jacobi matrices.

\section{Matrix Sturm-Liouville operators}
\label{sec-MatrixSL}

Let us consider the matrix Sturm-Liouville operator:
$$
H\;=\;
-\partial_x\bigl(p\,\partial_x+q\bigr)\;+\;q^*\partial_x\,+\,v
\;,
$$
where $p$, $q$ and $v=v^*$ are continuous functions on $[0,1]$ into the $L\times L$ matrices and $p$ is continuously differentiable and positive (definite) with a uniform lower bound $p\geq c\,\one_L$ for some constant $c>0$. For many of the results below, less regularity of $p$, $q$ and $v$ is sufficient. In particular, piecewise continuity of $q$ and $v$ and piecewise continuous differentiability of $p$ with finitely many pieces (as well as singular Kronig-Penney-like potentials) can be dealt with by working with boundary conditions at the discontinuities in a similar manner as described below. This is relevant for the analysis of Jacobi matrices later on, but we choose to avoid the associated technical issues in the first sections of the paper. Crucial is, however, the uniform lower bound on $p$. Vanishing of $p$ at the boundaries leads to a singular Sturm-Liouville operator with numerous interesting questions ({\it e.g.} Weyl extension theory) that are not dealt with here.  For now, $H$ will be considered as acting on all functions in the Sobolev space $H^2((0,1),\CM^{L})$, namely as the so-called maximal operator. Let us consider the Schr\"odinger equation $H\phi=E\phi$ at energy $E\in\RM$ which is a second order differential equation. It is known at least since Bott's seminal work \cite{Bot} that the standard rewriting of this second order linear equation as a first order equation leads to a special type of a Hamiltonian system. Indeed, let us set
\begin{equation}
\label{eq-identify}
\Phi(x)\;=\;
\left(\begin{array}{c}
\phi(x) \\
\big((p\,\partial_x+q)\phi \big)(x) 
\end{array}
\right)
\;,
\qquad
\Vv(x)\;=\;
\left(\begin{array}{cc}
\left(v-q^*p^{-1}q\right)(x) & \left(q^*p^{-1}\right)(x)\\
\left(p^{-1}q\right)(x) & \left(-p^{-1}\right)(x)
\end{array}
\right)
\;.
\end{equation}
Then $H\phi=E\phi$ is equivalent to
\begin{equation}
\label{eq-HamSys}
\bigl(\Jj\,\partial_x\,+\,\Vv(x)\bigr)\,\Phi(x)\;=\;E\;\Pp\,\Phi(x)\;,
\qquad
\Phi\in H^1((0,1),\CM^{2L})
\;,
\end{equation}
where 
\begin{equation}
\label{eq-identify2}
\Jj\;=\;
\begin{pmatrix}
0 & -\one_L \\ \one_L & 0
\end{pmatrix}
\;,
\qquad
\Pp\;=\;
\begin{pmatrix}
\one_L & 0 \\ 0 & 0
\end{pmatrix}
\;.
\end{equation}

Next let us recall that functions in $\phi\in H^2((0,1),\CM^{L})$ have limit values $\phi(0)$ and $(\partial_x\phi)(0)$, and similarly at $x=1$. Then one has for $\phi,\psi\in H^2((0,1),\CM^{L})$
\begin{equation}
\label{eq-SA}
\langle \phi\,|\,H\,\psi\rangle\,-\,\langle H\,\phi\,|\,\psi\rangle
\;=\;
\Phi(1)^*\Jj\Psi(1)\,-\,\Phi(0)^*\Jj\Psi(0)\;,
\end{equation}
where the scalar product on the l.h.s. is taken in $L^2((0,1),\CM^{L})$ and $\Phi$, $\Psi$ on the r.h.s. are associated to $\phi,\psi$ as in \eqref{eq-identify}. 

\vspace{.2cm}

Selfadjont boundary conditions now have to assure that the r.h.s. of \eqref{eq-SA} vanishes. Here the focus will be on separate boundary conditions specified by two $\Jj$-Lagrangian planes at the boundary points $0$ and $1$. Recall that a $\Jj$-Lagrangian plane is an $L$-dimensional subspace of $\CM^{2L}$ on which $\Jj$ vanishes as a quadratic form. Such an $L$-dimensional subspace will here always be given as the range of a matrix $\Psi\in\CM^{2L\times L}$ of full rank $L$ and satisfying 
\begin{equation}
\label{eq-Lagrange}
\Psi^*\Jj\Psi
\;=\;
0
\;.
\end{equation}
Note that two such matrices $\Psi$ and $\Phi$ specify the same $\Jj$-Lagrangian plane if and only if there is an invertible matrix $c\in\CM^{L\times L}$ such that $\Psi=\Phi c$. In this case, we say that $\Psi$ and $\Phi$ are equivalent and denote this by $\Psi\sim\Phi$. Note that this indeed defines an equivalence relation on the space of matrices in $\Psi\in \CM^{2L\times L}$ of full rank $L$ satisfying \eqref{eq-Lagrange}. The set of equivalence classes is denoted by 
$$\LM_L
\;=\;
\{[\Psi]_\sim:\Psi\in \CM^{2L\times L} \mbox{ of full rank }L\mbox{ and }\Psi^*\Jj\Psi=0\}
\;,
$$ 
and called the Lagrangian Grassmannian. 

\vspace{.2cm}

Now let $[\Psi_0]_\sim,[\Psi_1]_\sim\in\LM_L$ and define the following domain for $H$:
\begin{equation}
\label{eq-Domain}
\Dd_{\Psi_0,\Psi_1}(H)
\;=\;
\left\{
\phi\in H^2((0,1),\CM^{L})
\;:\;
\Phi(j)\in\Ran(\Psi_j)\;,\;j=0,1
\right\}
\;.
\end{equation}
Note that this indeed only depends on the classes $[\Psi_0]_\sim$ and $[\Psi_1]_\sim$. The conditions $\Phi(j)\in\Ran(\Psi_j)$ assure that  both terms on the r.h.s. of \eqref{eq-SA} vanish, and not only their difference (periodic boundary conditions are of a different type, but can be analyzed similarly \cite{SB2}). Therefore $H$ restricted to $\Dd_{\Psi_0,\Psi_1}(H)$ is a selfadjoint operator, which is denoted by $H_{\Psi_0,\Psi_1}$. Dirichlet boundary conditions at the left and right boundary correspond to the choices $\Psi_0=\Psi_1=\Psi_D$
$$
\Psi_D
\;=\;
\begin{pmatrix}
0 \\ \one_L 
\end{pmatrix}
\;.
$$
It is, moreover, a standard result that the selfadjoint operator $H_{\Psi_0,\Psi_1}$  has a compact resolvent so that it has discrete real spectrum. These eigenvalues can be calculated by looking for solutions of the Schr\"odinger equation $H\phi=E\phi$ in the domain $\Dd_{\Psi_0,\Psi_1}(H)$. Of course, any other finite interval instead of $[0,1]$ can be considered in the same manner, and it is also possible to work with periodic boundary conditions. For sake of concreteness, we restrict to the case described above.

\section{Hamiltonian systems}
\label{sec-HamSys}

The fundamental solution $\Tt^E(x)$ of \eqref{eq-HamSys} is given by
\begin{equation}
\label{eq-fundamental}
\partial_x \Tt^E(x)\;=\;
\Jj\,\bigl(\Vv(x)\,-\,E\,\Pp\bigr)\,\Tt^E(x)\;,
\qquad
\Tt^E(0)\;=\;\one_{2L}
\;.
\end{equation}
This is a particular case of a Hamiltonian system of the form
\begin{equation}
\label{eq-HamSysGen}
\partial_x \Tt(x)\;=\;
\Jj\,\Hh(x)\,\Tt(x)\;,
\qquad
\Tt(0)\;=\;\one_{2L}
\;,
\end{equation}
where $\Hh(x)$ is continuous and pointwise selfadjoint $\Hh(x)^*=\Hh(x)$. It is called the classical Hamiltonian. To recover the special case \eqref{eq-fundamental}, one chooses $\Hh(x)$ to be
\begin{equation}
\label{eq-HamEnergy}
\Hh^E(x)\;=\;\Vv(x)-E\,\Pp
\;.
\end{equation}
with $\Pp$ independent of $x$ and given by \eqref{eq-identify2}. The focus will be on this case stemming from a matrix Sturm-Liouville operator and in this case the fundamental solution will be denoted by $\Tt^E(x)$ instead of simply $\Tt(x)$. However, some results also hold for the general Hamiltonian system \eqref{eq-HamSysGen} and other Hamiltonian systems depending on an energy parameter  as in \eqref{eq-HamEnergy} with general positive $\Pp(x)$.  As the following example shows, such systems can be of interest.

\vspace{.2cm}

\noindent {\bf Example} $\;$ If $\Vv(x)=\Vv(x)^*$ is an arbitrary continuous matrix-valued function, not necessarily of the form given in \eqref{eq-identify}, the l.h.s. of \eqref{eq-HamSys} is given in terms of a one-dimensional Dirac-type operator $D=\Jj\,\partial_x+\Vv(x)$. If one furthermore chooses $\Pp=\one_{2L}$, then \eqref{eq-HamSysGen} is simply the associated eigenvalue equation $D\Phi=E\Phi$ if the energy dependent classical Hamiltonian is \eqref{eq-HamEnergy}. One also needs selfadjoint boundary conditions. As
$$
\langle \Phi\,|\,D\,\Psi\rangle\,-\,\langle D\,\Phi\,|\,\Psi\rangle
\;=\;
\Phi(1)^*\Jj\Psi(1)\,-\,\Phi(0)^*\Jj\Psi(0)\;,
\qquad
\Phi,\Psi\in H^1((0,1),\CM^{2L})
\;,
$$
and if one focuses again on separate boundary conditions, they are again given by two $\Jj$-Lagrangian planes as in \eqref{eq-Lagrange}. This allows to define a selfadjoint operator $D_{\Psi_0,\Psi_1}$ with domain $\Dd_{\Psi_0,\Psi_1}(D)$ as in \eqref{eq-Domain}.
\hfill $\diamond$

\vspace{.2cm}

It turns out that the positivity property
\begin{equation}
\label{eq-HamSysPos}
-\,\binom{0}{\one}^*\,\Hh(x)\,\binom{0}{\one}
\;>\;0
\;,
\end{equation}
is crucial for the space oscillations analyzed in Section~\ref{sec-SpaceOsci}. For the matrix Sturm-Liouville case with \eqref{eq-HamEnergy} and $\Vv(x)$ and $\Pp$ as given in \eqref{eq-identify} and \eqref{eq-identify2} respectively, this holds for all $E\in\RM$ because the l.h.s. of \eqref{eq-HamSysPos} is equal to $p(x)^{-1}$ which is positive. On the other hand, the eigenvalue calculation by intersection theory (Theorem~\ref{theo-osci1}) and the positivity in the energy variable (Theorem~\ref{theo-osci2}) hold for arbitrary Hamiltonian systems \eqref{eq-HamSysGen} with $\Hh^E(x)=\Vv(x)-E\Pp(x)$ and $\Pp(x)> 0$, namely $\Pp(x)$ need not be constant for these results and given by \eqref{eq-identify2} nor is it necessary that \eqref{eq-HamSysPos} holds.

\section{Matrix Pr\"ufer phase and intersection theory}
\label{sec-MatrixPruefer}

The solution to \eqref{eq-HamSysGen} lies in the group $\Gg(L)=\{\Tt\in \CM^{2L\times 2L}\,:\,\Tt^*\Jj\Tt=\Jj\}$ which via the Cayley transform is isomorphic to the generalized Lorentz group U$(L,L)$ of inertia $(L,L)$. For a given initial condition $\Psi_0\in\CM^{2L\times L}$ of rank $L$ and satisfying \eqref{eq-Lagrange}, one then obtains a path 
$$
x\in [0,1]\;\mapsto\; \Phi(x)\;=\; \Tt(x){\Phi}_0
$$ 
of matrices spanning Lagrangian planes, namely $\Phi(x)$ satisfies $\Phi(x)^*\Jj \Phi(x)=0$ and is of rank $L$ so that $[\Phi(x)]_\sim\in\LM_L$. If the Hamiltonian depends on $E$, then so does $\Tt^E(x)$ and thus also $\Phi^E(x)$ carries an upper index $E$. This path leads to an eigenfunction of the operator $H_{\Psi_0,\Psi_1}$ (or $D_{\Psi_0,\Psi_1}$ if the example of the Dirac operator is considered) for the eigenvalue $E$ if and only if the intersection of the planes spanned by $\Phi^E(1)$ and the right boundary condition $\Psi_1$ is non-trivial. More precisely, the dimension of this intersection is equal to the multiplicity of $E$ as an eigenvalue of $H_{\Psi_0,\Psi_1}$ (or $D_{\Psi_0,\Psi_1}$). If this intersection is non-trivial, one calls $1$ a conjugate point for the solution. More generally, given the above path $x\mapsto \Phi(x)$ and a fixed Lagrangian plane $[\Psi_1]_\sim$, one calls a point $x$ a conjugate point for the Hamiltonian system \eqref{eq-HamSysGen} if the intersection of (the span of) $\Phi(x)$ and $\Psi_1$ is non-trivial. The dimension of the intersection is called the multiplicity of the conjugate point. 

\vspace{.2cm}

The theory of intersections of Lagrangian planes is precisely described by the Bott-Maslov index. Most conveniently, it can be studied using the stereographic projection $\Pi:\LM_L\to \mbox{\rm U}(L)$ which is a real analytic bijection \cite{SB} that is (well-)defined by
$$
\Pi([\Phi]_\sim)
\;=\;
\begin{pmatrix} \one_L \\ \imath\one_L
\end{pmatrix}^*\Phi
\Big[\begin{pmatrix} \one_L \\ -\imath\one_L
\end{pmatrix}^*\Phi\Big]^{-1}
\;.
$$
Note, in particular, that the r.h.s. does not depend on the choice of the representative of the class $[\Phi]_\sim$. To shorten notation, we will also write 
$$
\Pi(\Phi)
\;=\;
\Pi([\Phi]_\sim)
\;.
$$ 
It is well-known ({\it e.g.} \cite{SB2}) that the dimension of the intersection of two $\Jj$-Lagrangian subspaces spanned by matrices $\Phi$ and $\Psi$ respectively is equal to the multiplicity of $1$ as an eigenvalue of the unitary $\Pi(\Psi)^*\Pi(\Phi)$. Furthermore, the Bott-Maslov index of a given (continuous) path $x\in [0,1]\mapsto [\Phi(x)]_\sim$ of $\Jj$-Lagrangian subspaces w.r.t. the singular cycle given by a Lagrangian subspace $[\Psi]_\sim$ is given by adding up all intersections with their multiplicity and orientation which is precisely given by the spectral flow of the path of unitaries $x\in [0,1]\mapsto \Pi(\Psi)^*\Pi(\Phi(x))$ through $1$. Intuitively, this counts the number of eigenvalues passing through $1$ in the positive sense, minus those passing in a negative sense. The spectral flow of a path is denoted by $\SF$, a notation that is used below. A particularly simple functional analytic definition of spectral flow is given by Phillips \cite{Phi}. All this is also described in detail in \cite{SB,SB2}. 

\vspace{.2cm}

For all the above reasons, it is reasonable to define the matrix Pr\"ufer phase by
\begin{equation}
\label{eq-MatrixPruefer}
U(x)\;=\;\Pi\bigl(\Tt(x){\Phi}_0\bigr)
\;.
\end{equation}
If the classical Hamiltonian $\Hh^E(x)$ depends on $E$, also $U^E(x)$  has an index to indicate this dependence. Then the above proves ({\it e.g.} \cite{SB,SB2}, but this is essentially known since the works of Bott and Lidski \cite{Bot,Lid}):

\begin{theo} 
\label{theo-osci1} The multiplicity of $x$ as conjugate point w.r.t. $\Psi_1$ is equal  to the multiplicity of $1$ as eigenvalue of the unitary $\Pi(\Psi_1)^*U(x)$. For a matrix Sturm-Liouville operator $H_{\Psi_0,\Psi_1}$, the multiplicity of $E$ as an eigenvalue of $H_{\Psi_0,\Psi_1}$ is equal to the multiplicity of $1$ as eigenvalue of the unitary $\Pi(\Psi_1)^*U^E(1)$.  
\end{theo}

Let us note that for Dirichlet boundary condition at $x=1$, one has $\Pi(\Psi_1)=\Pi(\Psi_D)=-\one$ so that one is interested in the eigenvalue $-1$ of the Pr\"ufer matrix $U^E(1)$.

\section{Positivity of Pr\"ufer phases in the energy variable}
\label{sec-EnergyOsci}

The next result states a crucial positivity property intrinsic to Hamiltonian systems with classical Hamiltonian $\Hh^E(x)=\Vv(x)-E\Pp(x)$ with $\Pp(x)\geq 0$. It dates back to Bott \cite{Bot} and the proof is reproduced from \cite{SB2} for the convenience of the reader and because it serves as a preparation for the arguments following further down.

\begin{theo} 
\label{theo-osci2} 
Consider the matrix Pr\"ufer phase $U^E(x)$ defined by \eqref{eq-MatrixPruefer} associated with the fundamental solution of \eqref{eq-HamSysGen} for a classical Hamiltonian $\Hh^E(x)=\Vv(x)-E\Pp(x)$ with $\Pp(x)\geq 0$. For all $x\in(0,1]$, one has
\begin{equation}
\label{eq-UERot}
\frac{1}{\imath}(U^E(x))^*\partial_EU^E(x)
\;\geq \;
0
\;.
\end{equation}
As a function of $E$, the eigenvalues of ${U}^{E}(x)$ rotate around the unit circle in the positive sense. 

\vspace{.1cm}

\noindent If $\Vv(x)$ and $\Pp$ are given by \eqref{eq-identify} and \eqref{eq-identify2} respectively, then the inequality in \eqref{eq-UERot} is strict.
\end{theo}

\noindent {\bf Proof.} 
Let us introduce
$\phi^E_\pm(x)=(\one\,\pm\imath\one)\,{\Phi}^E(x)$ where ${\Phi}^E(x)=\Tt^E(x){\Phi}_0$. Then ${\Phi}^E(x)$ span $\Jj$-Lagrangian subspaces and thus $\phi^E_\pm(x)$ are invertible $L\times L$ matrices.  One has by definition
$$
{U}^E(x)
\;=\;
\phi_-^E(x)(\phi_+^E(x))^{-1}
\;=\;
((\phi_-^E(x))^{-1})^*(\phi_+^E(x))^*
\;.
$$ 
Now
\begin{align*}
\frac{1}{\imath}\;
{U}^E(x)^*\,\partial_E\,{U}^E(x)
&
\;=\;
((\phi_+^E(x))^{-1})^*
\,\frac{1}{\imath}\,
\Bigl[\,
(\phi_-^E(x))^*\partial_E\phi_-^E(x) \,-\,
(\phi_+^E(x))^*\partial_E\phi_+^E(x)
\,\Bigr]
(\phi_+^E(x))^{-1}
\\
&
\;=\;
((\phi_+^E(x))^{-1})^*
\,2\,
({\Phi}^E(x))^*\,{\Jj}\,\partial_E {\Phi}^E(x)
(\phi_+^E(x))^{-1}
\\
&
\;=\;
2\,(\Psi_0(\phi_+^E(x))^{-1})^*
\,\Tt^E(x)^*\,{\Jj}\,\partial_E \Tt^E(x)\,\Psi_0
(\phi_+^E(x))^{-1}
\;.
\end{align*}
Thus it is sufficient to verify the positive definiteness  $\Tt^E(x)^*\Jj\partial_E\Tt^E(x)\geq 0$. For that purpose, let $\epsilon>0$. By \eqref{eq-fundamental},
$$
\partial_y\,\left(\Tt^E(y)^*\,\Jj\,\Tt^{E+\epsilon}(y)\right)
\;=\;
\epsilon\;\Tt^E(y)^*\,\Pp(y)\,\Tt^{E+\epsilon}(y)
\;.
$$
As $\Tt^E(x)^*\,\Jj\,\Tt^{E}(x)=\Jj=\Tt^E(0)^*\,\Jj\,\Tt^{E+\epsilon}(0)$, one thus has
\begin{eqnarray}
\Tt^E(x)^*\Jj\partial_E\Tt^E(x)
& = & 
\lim_{\epsilon\to 0}\;
\epsilon^{-1}\,
\left(\Tt^E(x)^*\,\Jj\,\Tt^{E+\epsilon}(x)\,-\,
\Tt^E(x)^*\,\Jj\,\Tt^{E}(x)
\right)
\nonumber
\\
& = & 
\lim_{\epsilon\to 0}\;
\epsilon^{-1}\,
\left(\Tt^E(x)^*\,\Jj\,\Tt^{E+\epsilon}(x)\,-\,
\Tt^E(0)^*\,\Jj\,\Tt^{E+\epsilon}(0)
\right)
\nonumber
\\
& = &
\lim_{\epsilon\to 0}\;
\int^x_0dy\;\Tt^E(y)^*\,\Pp(y)\,\Tt^{E+\epsilon}(y)
\nonumber
\\
& = &
\int^x_0dy\;\Tt^E(y)^*\,\Pp(y)\,\Tt^E(y)
\;.
\label{eq-PosMatrix}
\end{eqnarray}
Because $\Pp(y)$ is non-negative, this implies the claim \eqref{eq-UERot}. The second statement follows from first order perturbation theory \cite{Kat}. For the proof of the final statement, it is sufficient to show that the integrand $\Tt^E(y)^*\,\Pp(y)\,\Tt^E(y)$ is strictly positive for $y$ sufficiently small. Indeed, it follows from \eqref{eq-fundamental} that $\Tt^E(y)= \one+y\, \Jj^* (E\,\Pp-\Vv(y))+\Oo(y^2)$. Thus replacing \eqref{eq-identify} and \eqref{eq-identify2} shows
$$
\Tt^E(y)^*\,\Pp\,\Tt^E(y)
\;=\;
\begin{pmatrix}
\one & 0 \\ 0 & 0
\end{pmatrix}
\;-\;y\,
\begin{pmatrix}
q^*p^{-1}+p^{-1}q & -p^{-1} \\ -p^{-1} & 0
\end{pmatrix}
\;+\;\Oo(y^2)
\;.
$$
For $y$ sufficiently small, this is indeed a strictly positive matrix.
\hfill $\Box$

\vspace{.2cm}

As all intersections  of the path $E\mapsto \Pi(\Psi_1)^*U^E(1)$ are in the positive sense by Theorem~\ref{theo-osci2}, one deduces the following result connecting the eigenvalue counting of $H_{\Psi_0,\Psi_1}$ to the Bott-Maslov index of that path:

\begin{coro} 
\label{coro-Eosci} One has
$$
\#\{\mbox{\rm eigenvalues of } H_{\Psi_0,\Psi_1}\;\leq\;E\}
\;=\;
\SF\big(e\in(-\infty,E]\mapsto \Pi(\Psi_1)^*U^e(1)\;\mbox{\rm through }1\big)
\;,
$$
where the spectral flow counts the number of unit eigenvalues passing through $1$ in the positive sense {\rm (}necessarily by {\rm Theorem~\ref{theo-osci2})}, counted with their multiplicity. 
\end{coro}

\section{Positivity of Pr\"ufer phases in the space variable}
\label{sec-SpaceOsci}

The following result now concerns residual positivity properties of the matrix Pr\"ufer variables in the spatial coordinate under the condition that \eqref{eq-HamSysPos} holds.  It is essentially a corollary of Theorem~V.6.2 of \cite{Atk}, but we provide a direct proof.

\begin{theo} 
\label{theo-osci3} 
Consider matrix Pr\"ufer phase \eqref{eq-MatrixPruefer} associated with the fundamental solution of the Hamiltonian system \eqref{eq-HamSysGen} with \eqref{eq-HamSysPos}.  For all $x\in(0,1)$, one has on the subspace $\Ker(U(x)+\one_L)$ 
$$
\frac{1}{\imath}(U(x))^*\partial_xU(x)\Big|_{\Ker(U(x)+\one_L)}
\;> \;
0
\;.
$$
As a function of $x$, the eigenvalues of ${U}(x)$ pass through $-1$ only in the positive sense.
\end{theo}

\noindent {\bf Proof.} The same objects as in the proof of Theorem~\ref{theo-osci2} will be used, but the index $E$ will be dropped and also the argument $x$ on $U(x)$, $\Phi(x)$ and $\phi_\pm(x)$. Also let us introduce the upper and lower entry of $\Phi$ as $\phi_0$ and $\phi_1$, namely $\phi_\pm=\phi_0\pm \imath\phi_1$. As in the proof of Theorem~\ref{theo-osci2}, one first checks that
$$
\frac{1}{\imath}\;(U)^*\,\partial_x\,U
\;=\;
2\,((\phi_+)^{-1})^*\Psi_0^*\,
(\Tt)^*\,{\Jj}\,\partial_x \Tt\,\Psi_0 (\phi_+)^{-1}
$$
Replacing the equation for the fundamental solution \eqref{eq-fundamental} thus gives 
\begin{align}
\frac{1}{\imath}\;(U)^*\,\partial_x\,U
& \;=\;
-\,2\,((\phi_+)^{-1})^*(\Psi_0)^*\,(\Tt)^*\,
\Hh
\,\Tt\,\Psi_0\,(\phi_+)^{-1}
\nonumber
\\
& 
\;=\;
-\,2\,((\phi_+)^{-1})^*(\Phi)^*\,
\Hh
\,\Phi\,(\phi_+)^{-1}
\label{eq-UPrime}
\end{align}
Now let $v\in\Ker(U+\one_L)$, namely $-v=Uv=((\phi_-)^{-1})^*(\phi_+)^*v$ or equivalently $-(\phi_-)^*v=(\phi_+)^*v$ or yet simply $(\phi_0)^*v=0$. But, as $(\phi_0)^*\phi_1=(\phi_1)^*\phi_0$ by the Lagrangian property of $\Phi$,
\begin{align*}
(\phi_0)^*v
&
\;=\;
(\phi_0)^*\phi_+(\phi_+)^{-1}v
\;=\;
(\phi_0)^*(\phi_0+\imath\phi_1)(\phi_+)^{-1}v
\;=\;
(\phi_-)^*\phi_0(\phi_+)^{-1}v
\;.
\end{align*}
Thus by the invertibility of $\phi_-$ one thus concludes
\begin{align*}
v\in\Ker(U+\one_L)
\quad
 \Longleftrightarrow\quad
\phi_0(\phi_+)^{-1}v\;=\;0
\quad \Longleftrightarrow\quad
\Phi\,(\phi_+)^{-1}v
\;=\;
\begin{pmatrix}
0 \\
w
\end{pmatrix}
\;,
\end{align*}
for some vector $w$.
Moreover, one checks $v\not=0$ if and only if $w\not =0$. Finally replacing in the above \eqref{eq-UPrime}, one finds for all $v\in\Ker(U+\one_L)$
\begin{align*}
v^*\,\frac{1}{\imath}\;(U)^*\,\partial_x\,U\, v
&
\;=\;
-\,2\,\begin{pmatrix}
0 \\
w\end{pmatrix}^*
\Hh
\begin{pmatrix}
0 \\w
\end{pmatrix}
\;.
\end{align*}
Thus \eqref{eq-HamSysPos} completes the proof of the claimed positivity. The last statement follows again from first order perturbation theory \cite{Kat}.
\hfill $\Box$

\section{Asymptotics and global properties of Pr\"ufer phase}
\label{sec-Asym}

Next let us examine the low-energy asymptotics of the matrix Pr\"ufer phases of a matrix Sturm-Liouville operator. Hence the classical Hamiltonian $\Hh^E(x)$ depends on $E$ with $\Pp$ as in \eqref{eq-identify2}. The outcome is the continuous analogue of results in \cite{GSV} (even though only less detailed information is provided here).

\begin{proposi} 
\label{prop-EnergyAsymp} 
For a matrix Sturm-Liouville operator, one has for any boundary condition $\Psi_0$ and any $x>0$,
$$
\lim_{E\to -\infty}\;U^E(x)
\;=\;
-\,\one
\;.
$$
Moreover, if $\Psi_0\cap\Psi_D=\{0\}$ and $0<x\leq C (-E)^{-1}$ for some constant $C>0$,
$$
\frac{1}{\imath}\;U^E(x)^*\,\partial_x\,U^E(x)
\;<\;0
\;.
$$
\end{proposi}

\noindent {\bf Proof.} 
For the analysis of the fundamental solution of \eqref{eq-fundamental} in the limit $E\to-\infty$, let us consider the rescaled object
$$
\widetilde{\Tt}^E(y)
\;=\;
\Tt^E(-E^{-1}y)
\;,
\qquad
y\in [0,-E]
\;.
$$
It satisfies
$$
\partial_y \widetilde{\Tt}^E(y)\;=\;
\Jj^*\,\bigl(\Pp\,-\,E^{-1}\Vv(-E^{-1}y)\bigr)\,\widetilde{\Tt}^E(y)\;,
\qquad
\widetilde{\Tt}^E(0)\;=\;\one_{2L}
\;.
$$
Thus
$$
\widetilde{\Tt}^E(y)\;=\;
\one_{2L}\;+\;
\int^y_0dz\;
\bigl(\Jj^*\Pp\,-\,E^{-1}\Jj^*\Vv(-E^{-1}z)\bigr)\,\widetilde{\Tt}^E(z)
\;.
$$
A Dyson series argument using $\|\Vv\|_\infty<C<\infty$ and the explicit form $\Jj^*\Pp$ thus shows
$$
\widetilde{\Tt}^E(y)
\;=\;
\begin{pmatrix}
\one & 0
\\
-y & \one
\end{pmatrix}
\;+\;
\Oo(|E|^{-1}y)
\;.
$$
Hence
\begin{equation}
\label{eq-TMatDevelop}
{\Tt}^E(x)
\;=\;
\begin{pmatrix}
\one & 0
\\ Ex & \one
\end{pmatrix}
\;+\;
\Oo(x)
\;,
\end{equation}
with an error term that is uniformly bounded in $E$. Hence using the matrix M\"obius transformation and $U^E(0)=\Pi(\Psi_0)$,
$$
U^E(x)
\;=\;
\Pi\big(
{\Tt}^E(x)\Psi_0\big)
\;=\;
\left(
\one\,-\,
\tfrac{\imath}{2}\, Ex
\begin{pmatrix}
\one & \one \\ -\one & -\one
\end{pmatrix}
\,+\,
\Oo(x)
\right)
\cdot
U^E(0)\;\longrightarrow\;-\one
\;,
$$
in the limit $E\to-\infty$ for $x>0$. 
The proof of the second claim is based on the identity \eqref{eq-UPrime}. Using \eqref{eq-TMatDevelop} let us thus evaluate
$$
(\Psi_0)^*\,(\Tt^E)^*\,
\bigl(E\,\Pp-\Vv\bigr)
\,\Tt^E\,\Psi_0
\;=\;
E\,(\Psi_0)^*\,\Pp\,\Psi_0
\;+\;\Oo(Ex)
\;.
$$
which already implies the claim because $\Psi_0\cap\Psi_D=\{0\}$ is equivalent to $(\Psi_0)^*\,\Pp\,\Psi_0>0$.
\hfill $\Box$

\begin{figure}
\centering
\includegraphics[width=8cm]{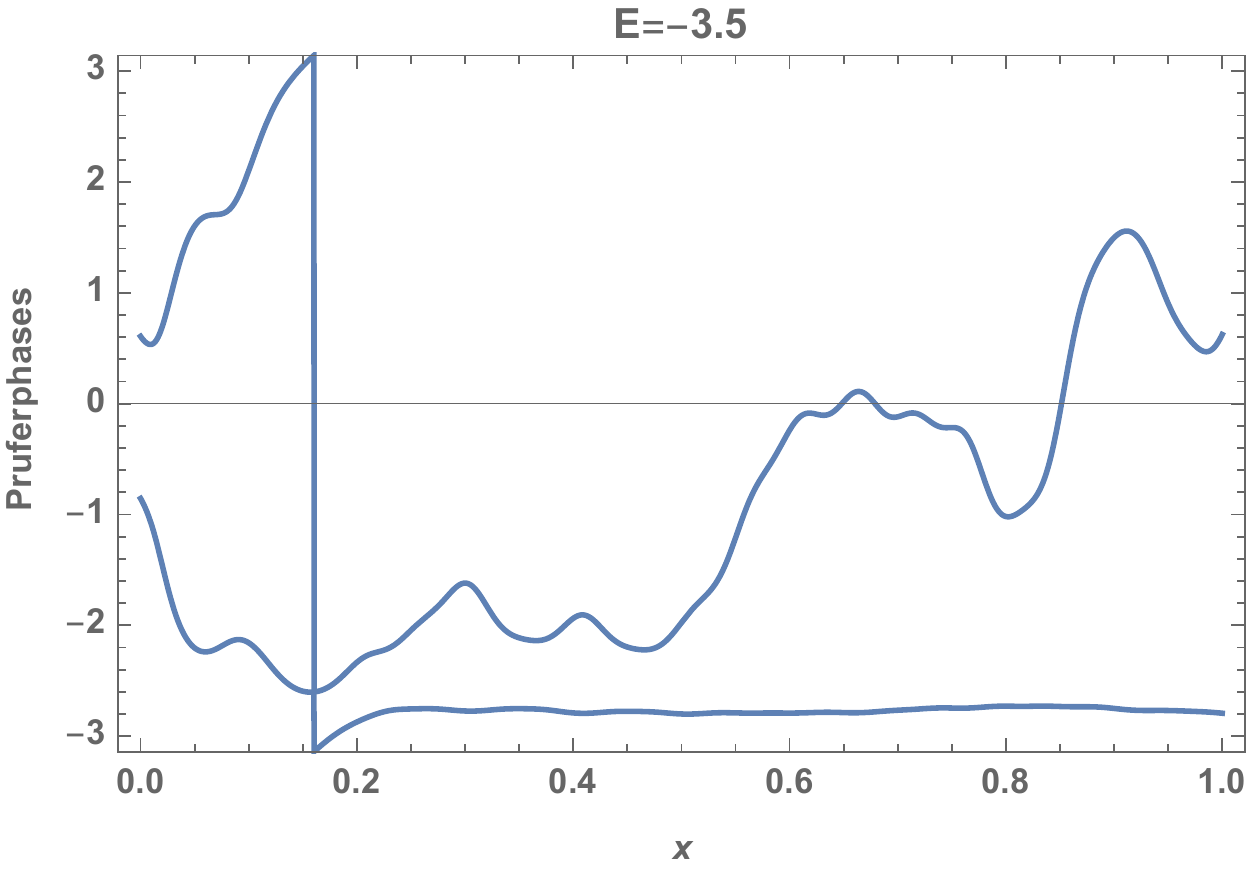} \hspace{.3cm}
\includegraphics[width=8cm]{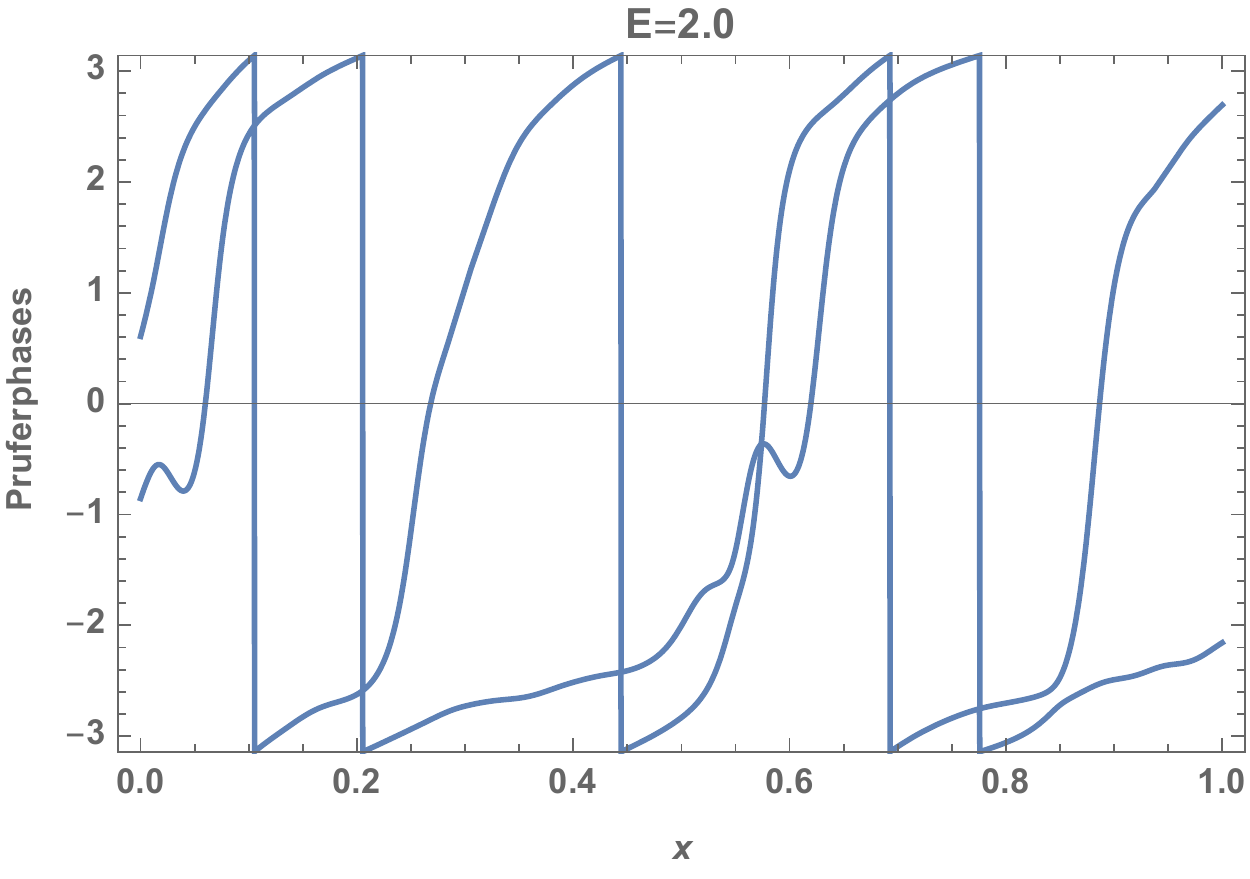}
\caption{The phases of the eigenvalues of $x\mapsto U^E(x)$ for the particular Sturm-Liouville operator described in Section~\ref{sec-Numerics} for two energies $E=-3.5$ and $E=2.0$ respectively. The vertical lines indicate a passage of one Pr\"ufer phase by $e^{\imath\pi}=-1$ and thus fix a conjugate point $x_c$ at which the Sturm-Liouville operator on $[0,x_c]$ with Dirichlet boundary condition at $x_c$ has an eigenvalue. The number of such points on $[0,1]$ is equal to the number of eigenvalues of $H_{\Psi_0,\Psi_D}$ below $E$. Hence there is one eigenvalue below $-3.5$ and five below $2.0$.}
\label{fig-SpaceOsci1}
\end{figure}

\begin{theo} 
\label{theo-osci4} 
For a matrix Sturm-Liouville operator with Dirichlet boundary condition at $x=1$,
$$
\#\{\mbox{\rm eigenvalues of } H_{\Psi_0,\Psi_D}\;\leq\;E\}
\;=\;
\SF\big(x\in[0,1]\mapsto U^E(x)\;\mbox{\rm through }-1\big)
\;,
$$
where the spectral flow counts the number of eigenvalues passing through $-1$ in the positive sense {\rm (}necessarily by {\rm Theorem~\ref{theo-osci3})}, counted with their multiplicity. 
\end{theo}

\noindent {\bf Proof.} 
By Proposition~\ref{prop-EnergyAsymp} there exists an $E_-$ such that for any $e\leq E_-$ the spectral flow of $x\in[0,1]\mapsto U^e(x)$ by $-1$ vanishes, namely there are no conjugate points in $[0,1]$ for all $e\leq E_-$. Furthermore, the spectral flow of $e\in(-\infty,E_-]\mapsto U^e(1)$ through $-1$ vanishes. Hence it is sufficient to consider the compactly defined continuous map $(x,e)\in[0,1]\times[E_-,E]\mapsto U^e(x)$. By the homotopy invariance, the spectral flow from $(0,E_-)$ to $(1,E)$ is independent of the choice of path. In particular, when one considers the spectral flow along the segments $[0,1]\times \{E_-\}$ and $1\times [E_-,E]$, it is equal to the number of eigenvalues of $ H_{\Psi_0,\Psi_D}$ below $E$ by Corollary~\ref{coro-Eosci}. On the other hand, let us consider the spectral flow along the segments $\{0\}\times [E_-,E]$ and $[0,1]\times \{E\}$. The spectral flow along $\{0\}\times [E_-,E]$ clearly vanishes as $U^e(0)$ is constant, and thus the second contribution leads to the statement. 
\hfill $\Box$

\section{Numerical illustration}
\label{sec-Numerics}

\begin{figure}
\centering
\includegraphics[width=8cm]{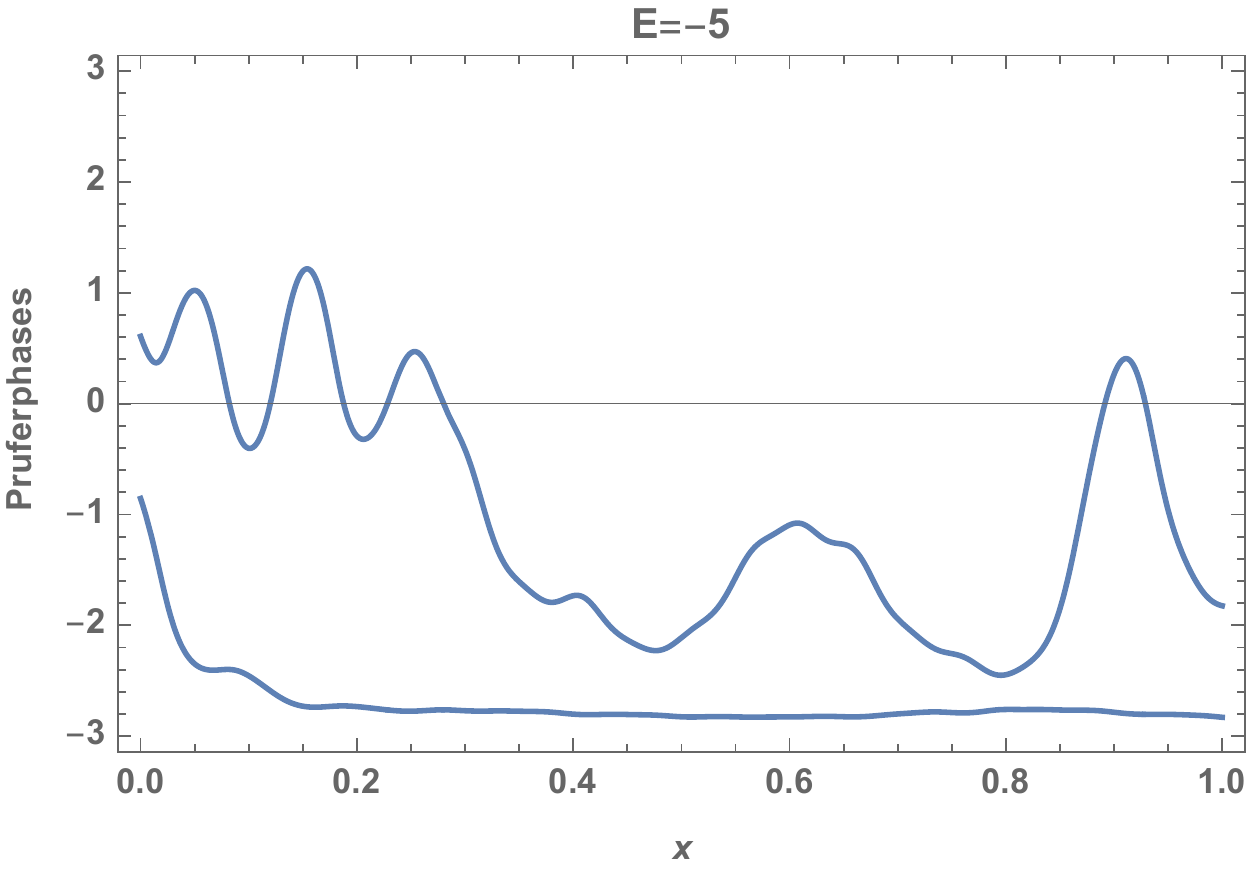} \hspace{.3cm}
\includegraphics[width=8cm]{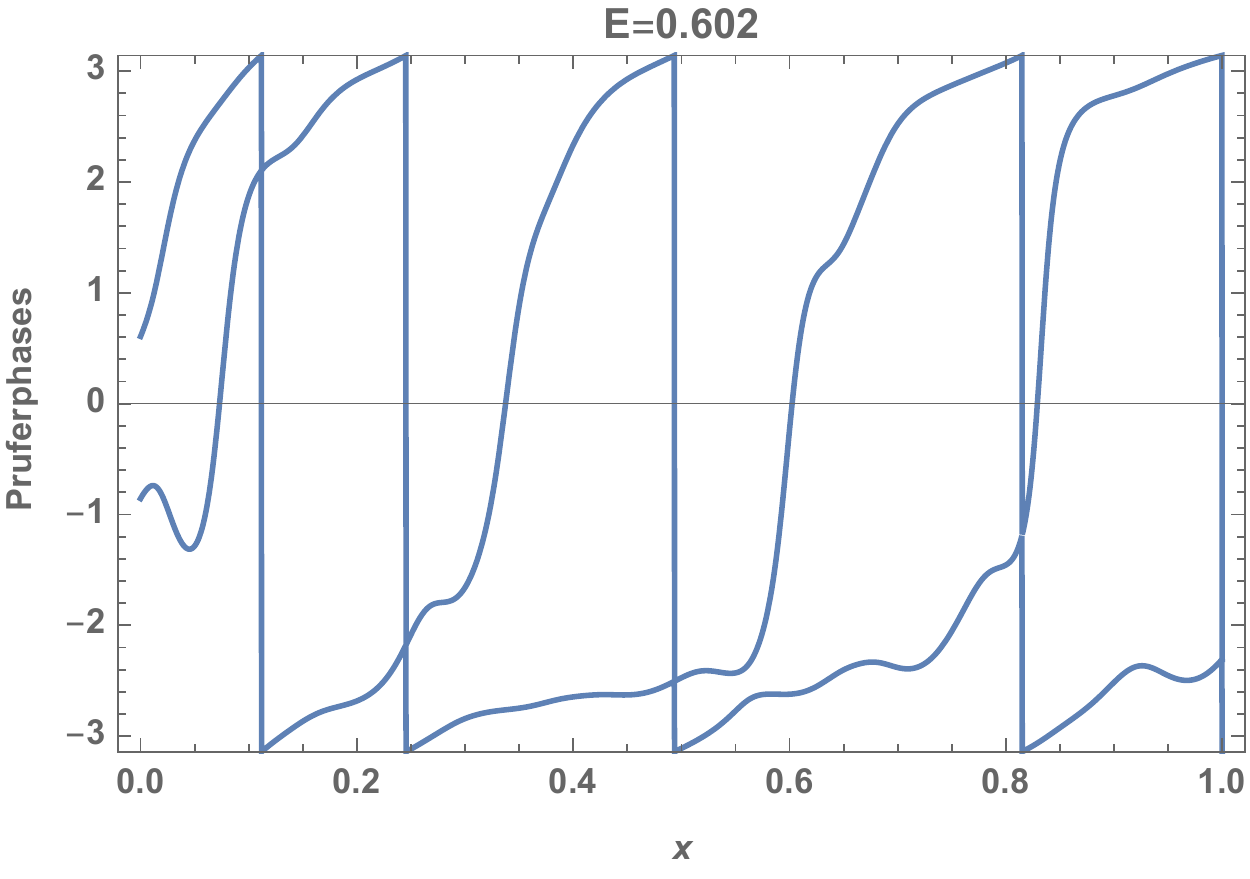}
\caption{These plots are the same as in Figure~1, but for two further energies.  Let us stress that the Pr\"ufer phases at $x=0$ are the same for all plots, and they are given by the (phases of the) eigenvalues of $\Pi(\Psi_0)$. The first plot of this figure is for energy $-5$ which (according to the plot) lies below the spectrum of $H_{\Psi_0,\Psi_D}$. Hence there is no passage of a Pr\"ufer phase by $-1$. This plot also illustrates Proposition~\ref{prop-EnergyAsymp}, namely the energy is already sufficiently small so that the eigenvalue slopes at $x=0$ are negative. The plot at $E=0.602$ is included because there is a passage by $-1$ of one of the two Pr\"ufer phases precisely at $x=1$. Therefore $E=0.602$ is an eigenvalue of $H_{\Psi_0,\Psi_D}$.}
\label{fig-SpaceOsci2}
\end{figure}

To illustrate the above results by a concrete example, we used a short Mathematica program that numerically solves for the matrix Pr\"ufer phase and its spectrum. Even though the particular form of matrix Sturm-Liouville operator may not be of great importance, let us spell it out explicitly anyhow. First of all, the fiber size is $L=2$ and the matrix valued coefficients were chosen (fairly randomly) to be
$$
p(x)
\;=\;
\begin{pmatrix}
2 + \cos(12 \,x) & \sin(11.5 \,x) \\ \sin(11.5 \,x)
&   3 - \sin(16 \,x) 
\end{pmatrix}
\;,
\qquad
q(x)\; =\; 
\begin{pmatrix}
3 & \cos(10 \,x)  
\\
0 & 3 \sin(20 \,x)
\end{pmatrix}
\;,
$$
and
$$
v(x)
\;=\;
\begin{pmatrix}
\cos(5\,x) & 7 \sin(61.5\, x) \\ 7 \sin(61.5\, x) & -2 + \sin(27.5\, x) 
\end{pmatrix}
\;.
$$
Finally, the left boundary condition is fixed to be
$$
\Psi_0 
\;=\;
\begin{pmatrix} M \\ \one_2 \end{pmatrix}
\;,
\qquad
M\;=\;
\begin{pmatrix}
2 & 1 \\ 
1 & -3 
\end{pmatrix}
\;.
$$
For a given energy $E\in\RM$, the fundamental equation \eqref{eq-fundamental} can be solved numerically and then allows to infer the matrix Pr\"ufer phase $U^E(x)$ via \eqref{eq-MatrixPruefer}. Its eigenvalues, namely the Pr\"ufer phases can then readily be calculated. Any of the plots shown in Figures 1-3 did not take longer than a few minutes on a laptop. The figure captions further discuss the outcome of the numerics in view of the results above.

\begin{figure}
\centering
\includegraphics[width=8cm]{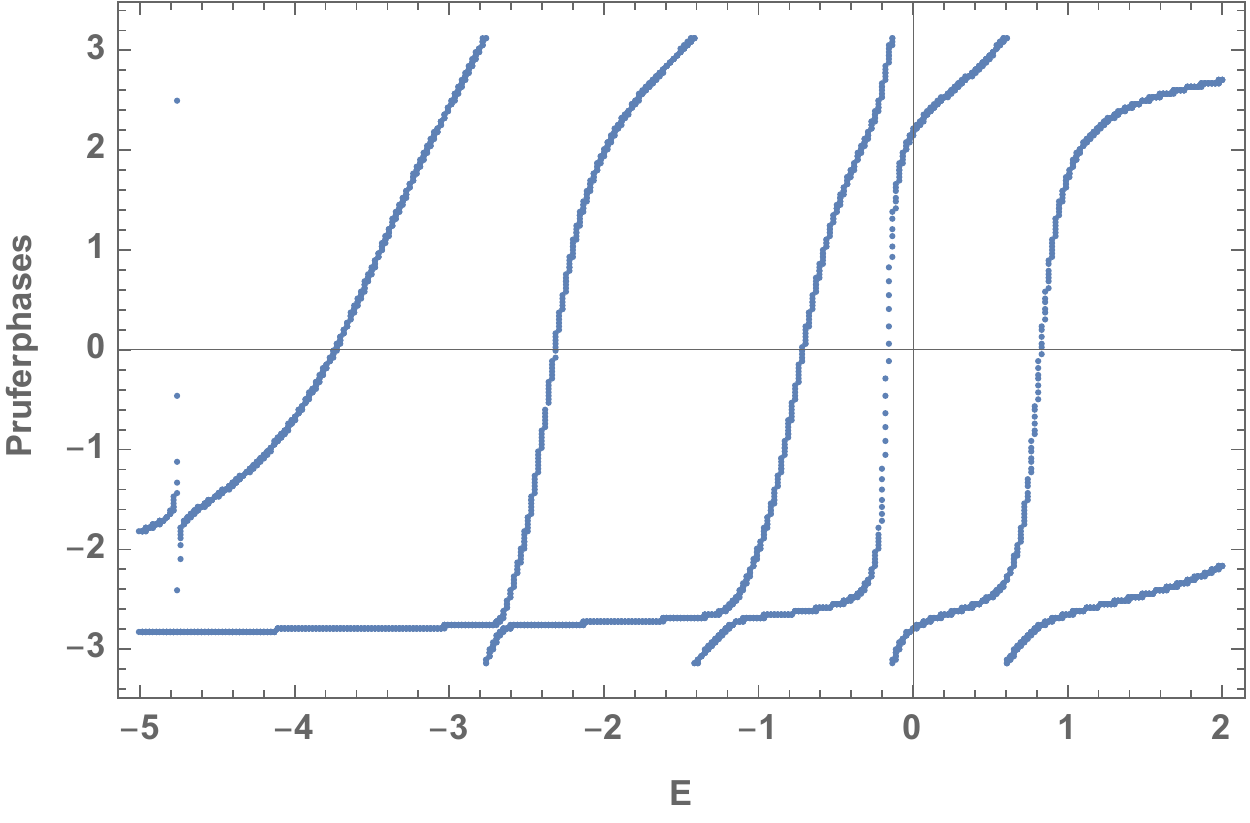} \hspace{.3cm}
\includegraphics[width=8cm]{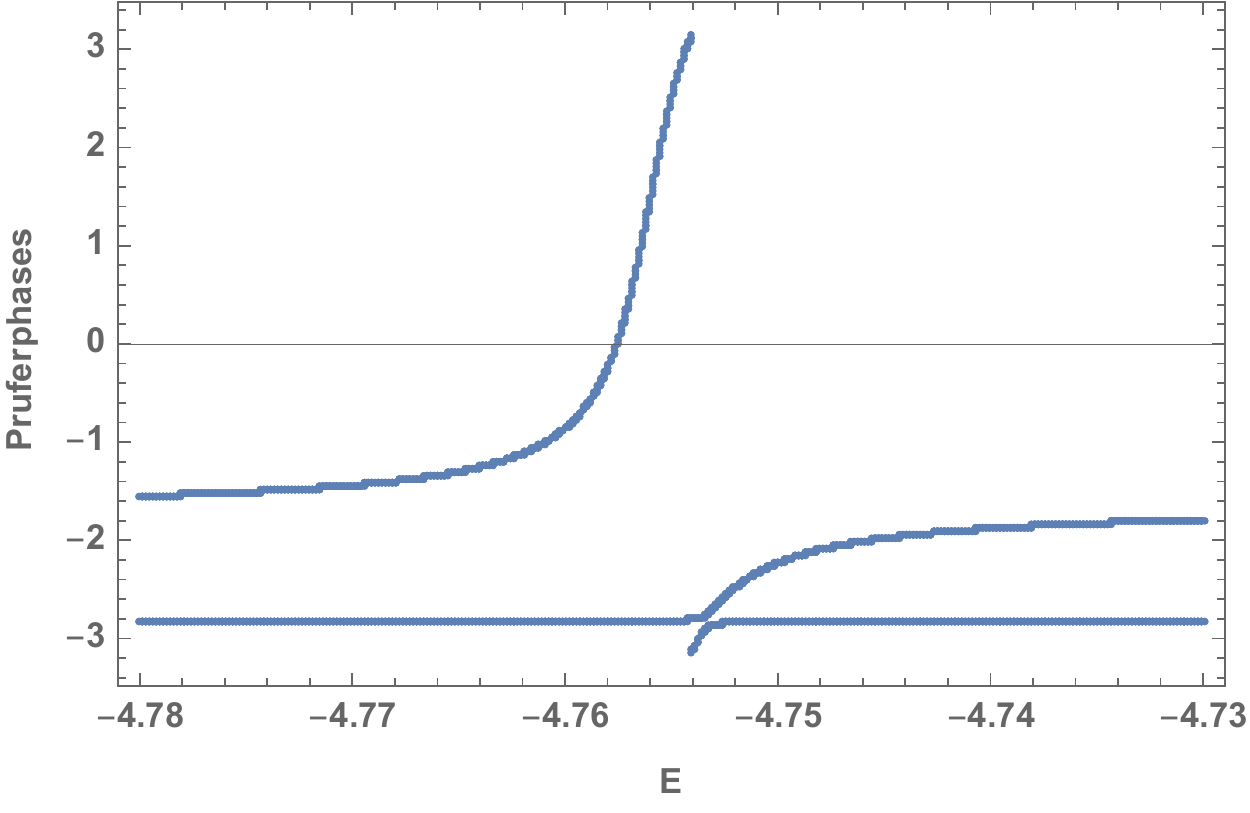}
\caption{The eigenvalues of $E\mapsto U^E(1)$  for the particular Sturm-Liouville operator described in Section~\ref{sec-Numerics}. For a discrete set of energies, the solution $x\in[0,1]\mapsto U^E(x)$ is calculated numerically to extend the matrix Pr\"ufer phase at $x=1$. One clearly observes the monotonicity of the Pr\"ufer phases in the energy variable. The eigenvalues of
$H_{\Psi_0,\Psi_D}$ are given by those energies at which one Pr\"ufer phase is equal to $-1$. The rough numerical analysis in the first figure may have missed the lowest eigenvalue at about $E=-4.766$ , but clearly the first plot of Figure~1 indicates that there must be one eigenvalue with energy less than $-3.5$, which is then readily found by the more careful numerical study in the second plot.
}
\label{fig-EnergyOsci}
\end{figure}

\section{Energy oscillations for matrix Jacobi operators}
\label{sec-Jacobi}

A matrix Jacobi operator of length $N\geq 3$ is a matrix of the form
\begin{equation}
\label{eq-matrix}
H_N
\;=\;
\left(
\begin{array}{ccccccc}
V_1       & T_2  &        &        &         &        \\
T_2^*      & V_2    &  T_3  &        &         &        \\
            & T_3^* & V_3    & \ddots &         &        \\
            &        & \ddots & \ddots & \ddots  &        \\
            &        &        & \ddots & V_{N-1} & T_N   \\
     &        &        &        & T_N^*  & V_N
\end{array}
\right)
\;,
\end{equation}
where $(V_n)_{n=1,\ldots,N}$ are selfadjoint complex $L\times L$ matrices and 
$(T_n)_{n=2,\ldots,N}$ are invertible complex $L\times L$ matrices.  The aim of the remaining part of the paper is to carry out a spectral analysis of $H_N$ by using suitably defined matrix Pr\"ufer phases and to discuss Sturm oscillation theory of these operators. This section reviews energy oscillations based essentially on \cite{SB}, then the remaining two sections provide two different approaches to study space oscillations of the matrix Pr\"ufer phases.

\vspace{.2cm}

To slightly simplify the set-up, let us start out with a gauge transformation (namely a strictly local unitary) denoted by  $G=\diag(G_1,\ldots,G_N)$ with $L\times L$ unitary matrices $G_n$, $n=1,\ldots,N$. Then
$$
GH_NG^*
\;=\;
\left(
\begin{array}{ccccccc}
G_1V_1G_1^*       & G_1T_2 G_2^*  &        &        &         &        \\
(G_1T_2G_2^*)^*      & G_2V_2G_2^*    & G_2 T_3 G_3^*  &        &         &        \\
            & (G_2T_3G_3^*)^* & G_3 V_3 G_3^*   & \ddots &         &        \\
            &        & \ddots & \ddots & \ddots  &        \\
            &        &        & \ddots & G_{N-1} V_{N-1}G_{N-1}^* & G_{N-1}T_N G_N^*  \\
     &        &        &        & (G_{N-1}T_NG_N^*)^*  & G_NV_N G_N^*
\end{array}
\right)
\;.
$$
Now one can iteratively choose the $G_n$. Start out with $G_1=\one$. Then choose $G_2$ to be the (unitary) phase in the polar decomposition of $T_2=G_2|T_2|$, next let $G_3$ be the phase of $G_2T_3=G_3|G_2T_3|$, and so on. One concludes that $GH_NG^*$ is again of the form of $H_N$ given in \eqref{eq-matrix}, but with positive off-diagonal terms. From now on, we thus suppose that $T_n>0$ for all $n=2,\ldots,N$. 

\vspace{.2cm}

Next let us introduce the $2L\times 2L$ transfer matrices $\Tt_n^E$ by
\begin{equation}
\label{eq-transfer}
\Tt_n^E
\;=\;
\left(
\begin{array}{cc}
(E\,{\bf 1}\,-\,V_n)\,T_n^{-1} & - T_n \\
T_n^{-1} & {\bf 0}
\end{array}
\right)
\;,
\qquad
n=1,\ldots,N
\;,
\end{equation}
with $T_1=\one$.  Then define $2L\times L$ matrices by
\begin{equation}
\label{eq-transferid}
\Phi^E_n
\;=\;
\Tt^E_n\,
\Phi^E_{n-1}
\;,
\qquad
n=1,\ldots,N
\;,
\end{equation}
and the initial condition 
\begin{equation}
\label{eq-JacDir}
\Phi^E_0
\;=\;
\begin{pmatrix}
\one \\ 0
\end{pmatrix}
\;,
\end{equation}
given by the left Dirichlet boundary condition. Of crucial importance is the conservation of the sesquilinear form
$$
\Jj
\;=\;
\begin{pmatrix} 0 & - \one \\ \one & 0
\end{pmatrix}
\;,
$$
namely $\Tt^E_n$ lies in the group 
$$
\Gg(L)
\;=\;
\big\{\Tt\in\CM^{2L\times 2L}\;:\;\Tt^*\Jj\Tt=\Jj
\big\}
\;.
$$ 
Moreover, $\Phi^E_n$ is $\Jj$-Lagrangian, namely its span is of dimension $L$ and $(\Phi^E_n)^*\Jj\Phi^E_n=0$. For each such $\Jj$-Lagrangian plane $\Phi$, one can define its stereographic projection $\Pi(\Phi)$, which is a unitary $L\times L$ matrix \cite{SB}. Finally let us introduce the matrix Pr\"ufer phases by
$$
U^E_n\;=\;\Pi(\Phi^E_n)\;.
$$

Now let us introduce $\phi^E_n\in\CM^{L\times L}$ for $n=0,1,\ldots,N+1$ as the matrix coefficients of
$$
\Phi^E_n
\;=\;
\left(
\begin{array}{c}
T_{n+1}\phi^E_{n+1} \\
\phi^E_n
\end{array}
\right)
\;.
$$
By definition, $\phi^E_0=0$ and $\phi^E_{1}=\one$. Furthermore, $\phi^E_{N+1}$ is associated to the point $N+1$ lying outside of the support $\{1,\ldots,N\}$. The matrix $\phi^E_{N+1}$ is, however, of great importance for the eigenvalue problem of $H_N$. More precisely, if $\phi^E_{N+1}=0$, the Schr\"odinger equation $H\phi^E=E\phi^E$ holds for $\phi^E=(\phi^E_n)_{n=1,\ldots,N}$. This is not typical, but if the intersection of $\Phi^E_N$ with the right boundary condition is non-trivial, namely there is a non-vanishing $v\in\CM^L$ such that
$$
\Phi^E_N\,v\;\in\;
\begin{pmatrix}
0 \\ \one
\end{pmatrix}
\,\CM^L
\;,
$$
then one can set
$$
\psi^E_n\;=\;\phi^E_n\,v
\;,
$$
which then defines $\psi^E=(\psi^E_n)_{n=1,\ldots,N}\in\CM^{LN}$ satisfying the Schr\"odinger equation
\begin{equation}
\label{eq-Schroedinger}
H_N\psi^E
\;=\;
E\,\psi^E\;.
\end{equation}
The dimension of the intersection of $\Phi^E_N$ with the right boundary condition can conveniently be calculated from intersection theory using the matrix Pr\"ufer phase $U^E_N$, namely the following statement analogous to Theorems~\ref{theo-osci1} and \ref{theo-osci2} holds \cite{SB}:

\begin{theo} 
\label{theo-Jacobi} 
The multiplicity of $E$ as eigenvalues $H_N$ is equal to the multiplicity of $-1$ as eigenvalue of $U^E_N$. Moreover,
$$
\frac{1}{\imath}\; (U_N^E)^*\partial_E U_N^E
\;>\;0
\;.
$$
As a function of energy $E$, the eigenvalues of ${U}^{E}_N$ rotate around the unit circle in the positive sense and with non-vanishing speed. Furthermore,
$$
\#\{\mbox{\rm eigenvalues of } H_N\;\leq\;E\}
\;=\;
\SF\big(e\in(-\infty,E]\mapsto U^e_N\;\mbox{\rm through }-1\big)
\;.
$$
\end{theo}

\section{Interpolating Pr\"ufer phases via Sturm oscillations}
\label{sec-PrueferJacobi}

To state an analogue of Theorem~\ref{theo-osci3} for matrix Jacobi operators is more delicate. Even for a one-dimensional fiber $L=1$ where the Sturm oscillation counts the number of sign changes of the wave function  (solution of the eigenvalue equation) along the discrete set $\{1,\ldots,N\}$, this requires some care as it is possible that the wave function has zeros. The surprisingly intricate analysis is carried out in \cite{Tes}. In order to deal with the matrix-valued case with $L>1$, a careful definition of a suitable path of unitaries interpolating between $U^E_{n-1}$ and $U^E_n$ is needed. One can then define the Sturm oscillation number of the Jacobi matrix as the intersection number of the interpolating path. In this section, the path is constructed using discrete Sturm oscillations which counts the number of sign changes of the principal solution $n\in\{1,\ldots,N\}\mapsto \Phi^E_n$. This theory is developed in a much more general set-up in the book \cite{DEH} and here we merely extract the information essential for the present purposes.

\vspace{.2cm}

Let us begin by introducing the matrix
\begin{equation}
\label{eq-SDef}
S^E_n
\;=\;
(\phi^E_n)^*
T_{n+1}\phi^E_{n+1}
\;.
\end{equation}
It is selfadjoint because
$$
(\Phi^E_n)^*\Jj\Phi^E_n\;=\;0
\qquad
\Longleftrightarrow
\qquad
\Big(
(\phi^E_n)^*
T_{n+1}\phi^E_{n+1} \Big)^*
\;=\;
(\phi^E_n)^*
T_{n+1}\phi^E_{n+1}
\;.
$$
Let us note that 
$$
S^E_1\;=\;E-V_1\;,
\qquad
S^E_2
\;=\;
(E-V_1)T_2^{-1}(E-V_2)T_2^{-1}(E-V_1)-(E-V_1)
\;,
$$
and that there is a recurrence relation
\begin{equation}
\label{eq-SIterate}
S^E_n
\;=\;
(\phi^E_n)^*(E-V_n)\phi^E_n\,-\,S^E_{n-1}
\;.
\end{equation}
For any $S=S^*\in\CM^{L\times L}$ let us recall the definition of the Morse index
$$
\Morse (S)
\;=\;
\Tr(\chi(S<0))
\;.
$$
In view of the definition \eqref{eq-SDef} of $S^E_n$, the index $\Morse (S^E_n)$ can be interpreted as the number of sign changes of the principal solution from site $n$ to $n+1$. It is the object of Sturm oscillation theory to connect the total number of sign changes to the eigenvalue counting. This is well-known to be a special case of oscillation theory for discrete symplectic systems, see \cite{DEH} for a detailed review of the history.  The following result and its proof condensate the arguments in \cite{DEH} and is thus, due to the particular set-up and the supplementary assumption on $E$ not being in a finite singular set, considerably shorter. For sake of notational convenience, let us also introduce the complement of the Morse index
$$
\MorseC  (S)
\;=\;
\Tr(\chi(S\geq 0))
\;=\;
L\,-\,\Morse (S)
\;.
$$
%

\begin{theo} 
\label{theo-DiscreteStrum} 
Suppose that $E$ is not in the finite singular set
$$
\Ss
\;=\;
\bigcup_{n=1,\ldots,N-1}\sigma(H_n)
\;.
$$
Then one has
$$
\#\{\mbox{\rm eigenvalues of } H_N\;\leq\;E\}
\;=\;
\sum_{n=1}^N
\,
\MorseC  (S^E_n)
\;.
$$
\end{theo}

\noindent {\bf Proof.} Let us set $N^E=\sum_{n=1}^N\,\MorseC  (S^E_n)$. The proof consists of showing that there is a subspace $\Ee^E_\leq\subset\CM^{NL}$ of dimension $N^E$ on which $H_N-E$ is non-positive definite, and a subspace $\Ee^E_>\subset\CM^{NL}$ of dimension $NL-N^E$ on which $H_N-E$ is positive definite. These subspaces will be constructed iteratively in $n$, that is for $*$ either $\leq $ or $>$
$$
\Ee^E_*
\;=\;
\bigoplus_{n=1}^N
\;\Ee^{E,n}_*
\;,
$$
with
$$
\Ee^{E,n}_*\,\subset\,\Ee^n\;,
$$
where
$$
\Ee^n\;=\;
\left\{
\psi=(\psi_1,\ldots,\psi_N)\in\CM^{NL}\;:\;
\psi_n\not=0\;\mbox{ and }\;\psi_{n+1}=\ldots=\psi_{N}=0
\right\}
\,\cup\,\{0\}
\;.
$$
By construction, these subspaces satisfy $\Ee^{E,n}_*\cap\Ee^{E,m}_*=\{0\}$ for $n\not=m$. 
Let us first construct $\Ee^{E,n}_> $. For this purpose, let us choose $v\in\CM^L$ such that
$$
v^*S^E_n v\,<\,0
\;.
$$
Writing out the definition of $S^E_n$, one then has $\phi^E_nv\not=0$. Setting
\begin{equation}
\label{eq-PsiConstruct}
\psi^{E,n}_{v}
\;=\;
\left(
\phi^E_1v,\ldots,\phi^E_nv,0,\ldots,0
\right)
\;,
\end{equation}
where $\phi^E$ is the principal solution constructed above, one thus has $\psi^{E,n}_{v}\in \Ee^n$. 

\vspace{.1cm}

Now $(H_N-E)\psi^{E,n}_{v}$ is supported only on the sites $n$ and $n+1$. This implies, first of all, that taking the scalar product with $\psi^{E,n}_{v}$, that is, multiplying from the left by $(\psi^{E,n}_{v})^*$, only the contribution at the site $n$ remains. Thus
$$
(\psi^{E,n}_{v})^*(H_N-E)\psi^{E,n}_{v}
\;=\;
-\,v^*(\phi^E_n)^*T_{n+1}\phi^E_{n+1}v
\;=\;
-\,v^*S^E_n v\;>\;0
\;.
$$
This can be done for all vectors $v$ satisfying $v^*S^E_n v< 0$. Therefore
$$
\dim(\Ee^{E,n}_>)
\;\geq\;
\Morse (S^E_n)
\;.
$$
Second of all, for all $k<n$, one has by construction and the above support property that
$$
\psi^*(H_N-E)\psi^{E,n}_{v}
\;=\;
0
\;,
\qquad
\psi\in \Ee^{k}
\;.
$$
This implies that
$$
\psi^*(H_N-E)\psi^{E,n}_{v}\;=\;0
\;,
\qquad
\psi\in
\bigoplus_{k=1}^{n-1}
\;\Ee^{E,k}_>
\;.
$$
Now let us argue inductively in $n$ and suppose that $H_N-E$ is positive definite on $\bigoplus_{k=1}^{n-1}
\;\Ee^{E,k}_>$. For $\psi\in\bigoplus_{k=1}^{n-1}
\;\Ee^{E,k}_>$ and all $\mu,\mu'\in\CM$ with either $\mu\not=0$ or $\mu'\not=0$,
$$
(\mu\psi +\mu'\psi^{E,n}_{v})^*(H_N-E)(\mu\psi +\mu'\psi^{E,n}_{v})
\;=\;
|\mu|^2\,\psi^*(H_N-E)\psi 
\;+\;
|\mu'|^2\,(\psi^{E,n}_{v})^*
(H_N-E)\psi^{E,n}_{v}
\;>\;0
\;,
$$
namely $H_N-E$ is positive definite on $\bigoplus_{k=1}^{n}\Ee^{E,k}_>$. 
Proceeding iteratively in $n$, one deduces that $H_N-E$ is positive definite on all $\Ee^{E}_>=\bigoplus_{n=1}^{N}\Ee^{E,n}_>$. Because the $\Ee^{E,n}_>$ have trivial intersection, it follows that
\begin{equation}
\label{eq-FirstBound}
\dim(\Ee^{E}_>)
\;\geq\;
\sum_{n=1}^N
\,
\dim(\Ee^{E,n}_>)
\;\geq\;
\sum_{n=1}^N
\,
\Morse (S^E_n)
\;=\;
NL\,-\,N^E
\;.
\end{equation}

Next let us construct the $\Ee^{E,n}_\leq$. Proceeding as above, let us work with 
vectors $v\in\CM^L$ satisfying
$$
v^*S^E_n v\,\geq \,0
\;,
$$
and construct $\psi^{E,n}_v$ as in \eqref{eq-PsiConstruct}. As before, if $v^*S^E_n v>0$, then $\phi^E_n v\not=0$. If
$v^*S^E_n v=0$, then one cannot conclude directly that $\phi^E_n v\not=0$. If, however, one would have $\phi^E_n v=0$, then $\psi^{E,n}_v$ restricted to the first $n-1$ sites is an eigenvector of $H_{n-1}$ with eigenvalue $E$, which is not possible for $E\not\in\Ss$. Thus again $\phi^E_n v\not=0$ and one can conclude $\dim(\Ee^{E,n}_\leq)\geq \MorseC  (S^E_n)$ and finish the argument as above, showing that
\begin{equation}
\label{eq-SecondBound}
\dim(\Ee^{E}_\leq)
\;\geq\;
\sum_{n=1}^N
\,
\MorseC  (S^E_n)
\;=\;
N^E
\;.
\end{equation}
Given the bounds \eqref{eq-FirstBound} and  \eqref{eq-SecondBound} combined with the fact that the subspaces $\Ee^{E}_\leq $ and $\Ee^{E}_>$ have trivial intersection, one concludes that $\Ee^{E}_\leq +\Ee^{E}_> $ has a dimension of at least $NL$. Therefore the two inequalities \eqref{eq-FirstBound} and  \eqref{eq-SecondBound} must be equalities and the claim follows.
\hfill $\Box$

\vspace{.2cm}

\noindent {\bf Remark} The main reason why the above proof is relatively short is the following: there are many subspaces on which $H_N-E$ is positive (or non-positive). This can be understood even in a two-dimensional situation with $N=2$ and $L=1$ for which $H_N-E$ is a $2\times 2$ matrix. If one of its eigenvalues is positive and one negative, then the set of positive vectors forms a bicone and all one-dimensional subspaces in this bicone are positive. This non-uniqueness leads to a lot freedom in the construction of these subspaces. The important point is that, nevertheless, the dimension of  all these subspaces allows to conclude how many positive eigenvalues $H_N-E$ must have. The same holds for the non-positive subspaces and as, moreover, the dimensions add up, one has fully determined the number of positive and non-positive eigenvalues of $H_N-E$.
\hfill $\diamond$

\vspace{.2cm}

\noindent {\bf Remark} If $E\in \Ss$ and say $E\in \sigma(H_{n-1})$ there is $v\in\CM^L$ such that $\phi^E v\not=0$ restricted to $\{1,\ldots,n-1\}$ is an eigenvector of $H_{n-1}$. Then $\phi^E_{n} v=0$. Moreover, $\phi^E_{n-1} v\not=0$ because otherwise the three-term recurrence relation would imply $\phi^Ev=0$. It follows from the definition that $S^E_{n} v=0$ and $S^E_{n-1} v=0$ (note that this also fits with \eqref{eq-SIterate}).  Hence, one faces the difficulty that at step $n$, one cannot add a new linearly independent vector to $\Ee^{E,n}_\leq$ for this vector $v\in\Ker(S^E_n)$. This issue is not addressed here and the reader is referred to \cite{DEH}.
\hfill $\diamond$

\vspace{.2cm}

Based on Theorem~\ref{theo-DiscreteStrum}, it is now possible to construct the desired paths $x\in [n-1,n]\mapsto W^{E}(x)$ of unitaries interpolating between $U^{E}_{n-1}$ and $U^{E}_n$ by setting
\begin{equation}
\label{eq-UConstruct}
W^E(x)
\;=\;
\left\{
\begin{array}{cc}
e^{-\imath\, 3(x-n+\frac{2}{3})\,Q^E_{n-1}} \;, & x\in[n-1,n-\tfrac{2}{3}]\;,
\\
e^{\imath \,3(x-n+\frac{2}{3})\,2\pi\,\chi(S^E_n\geq 0)}\;, & x\in[n-\tfrac{2}{3},n-\tfrac{1}{3}]\;,
\\
e^{\imath \,3(x-n+\frac{1}{3})\,Q^E_n} \;,& x\in[n-\tfrac{1}{3},n]\;,\;\;\;\;\;\;
\end{array}
\right.
\end{equation}
where $Q^E_n$ are defined using the principal branch $\mbox{\rm Log}$ of the logarithm as
$$
Q^E_n\;=\;
-\imath\,\mbox{\rm Log}(U^E_n)
\;.
$$
During the first and third parts of the path \eqref{eq-UConstruct}, $U^E_{n-1}$ and $U^E_n$ are deformed into the identity without any eigenvalue passing through $-1$, while in the middle part exactly $\MorseC(S^E_n)$ loops are inserted leading to a spectral flow through $-1$ equal to $\MorseC(S^E_n)$. Therefore Theorem~\ref{theo-DiscreteStrum} implies

\begin{coro} 
\label{coro-DiscreteStrum} 
Suppose that $E\not\in\Ss$ and that $W^E(x)$ is defined by \eqref{eq-UConstruct}. Then
\begin{equation}
\label{eq-SturmMatrixJac}
\#\{\mbox{\rm eigenvalues of } H_N\;\leq\;E\}
\;=\;
\SF\big(x\in[0,N]\mapsto W^E(x)\;\mbox{\rm through }-1\big)
\;.
\end{equation}
\end{coro}

One shortcoming of this result is that it excludes the finite set $\Ss$ of singular energies, another one that it is based on the somewhat artificial construction \eqref{eq-UConstruct} so that Corollary~\ref{coro-DiscreteStrum} is merely a restating of Theorem~\ref{theo-DiscreteStrum}. The following section provides another construction of the interpolations.

\section{Interpolating Pr\"ufer phases via Hamiltonian systems}
\label{sec-PrueferJacobi2}

This section provides an alternative approach to construct the interpolating paths $x\mapsto U^{E}(x)$ satisfying $U^{E}(x)=U^E_n$ for all $n=0,\ldots,N$. Moreover, the construction will be done continuously in $E$, however, only for energies below some critical energy $E_c$, or alternatively for all energies above some other critical energy (see the Remark below). These critical energies will be defined below and the restrictions in energy are imposed due to technical difficulties. The paths $x\mapsto U^{E}(x)$ themselves will be given in terms of the fundamental solution of a suitably constructed Sturm-Liouville operator (depending continuously on $E$) and the eigenvalues of $U^{E}(x)$ pass through $-1$ only in the positive direction. Hence, this section establishes a connection between matrix Jacobi operators and Sturm-Liouville operators. This is best done with a Sturm-Liouville operator having Dirichlet boundary conditions $\Psi_D$ both at the left and right boundary. To match this for the matrix Jacobi operator, let us add an artificial site $0$ with $T_0=\one$ and $V_0=0$ so that the left boundary condition is
\begin{equation}
\label{eq-0-step-hamiltonian}
\Phi_{-1}^E\;=\;\begin{pmatrix} 0 \\ \one \end{pmatrix}
\;=\;\Psi_D
\;.
\end{equation}

Once the continuous path $(x,E)\in[-1,N]\times [-\infty,E_c)\mapsto U^{E}(x)$ is constructed, one can again deduce a Sturm-Liouville-like oscillation in the spatial variable as in Corollary~\ref{coro-DiscreteStrum}. Indeed, one can use a homotopy argument on the square $[-1,N]\times[-\infty,E]$ because the contributions of the paths $x\in [-1,N]\mapsto U^{-\infty}(x)$ as well as $e\in[-\infty,E]\mapsto U^{e}(-1)$ vanish, so that the intersection number of $x\in [-1,N]\mapsto U^{E}(x)$ is equal to the intersection number of $e\in[-\infty,E]\mapsto U^{e}(N)$ which is known to be equal to the number of eigenvalues below $E$, see Theorem~\ref{theo-Jacobi}. Therefore
\begin{equation}
\label{eq-SturmMatrixJac2}
\#\{\mbox{\rm eigenvalues of } H_N\;\leq\;E\}
\;=\;
\SF\big(x\in[-1,N]\mapsto U^E(x)\;\mbox{\rm through }-1\big)
\;.
\end{equation}
Let us note that the piece $x\in[-1,0]\mapsto U^E(x)$ connects $U^E_{-1}=-\one$ to $U^E_0=\one$ and has no intersection with $-1$ so that one can also drop this piece in \eqref{eq-SturmMatrixJac2} which is hence the same statement as in Corollary~\ref{coro-DiscreteStrum}, albeit only for energies below $E_c$ and for different interpolating matrix Pr\"ufer phases. On the other hand, it is not necessary to exclude the set of critical energies. We expect that both approaches allow to prove \eqref{eq-SturmMatrixJac} for all energies, but this remains an open problem at this point.

\vspace{.2cm}

The procedure for the construction of the path $x\in [-1,N]\mapsto U^{E}(x)$ is the following: For each fixed $E<E_c$ and all $n=0,\ldots,N$, the results below allow to construct a selfadjoint $\Hh^E_n$ such that $\Tt_n^E=e^{\Jj\Hh_n^E}$. Moreover, it can be assured (due to the later choice of $E_c$) that each $\Hh^E_n$ satisfies the positivity property \eqref{eq-HamSysPos}. Then set
\begin{equation}\label{eq:interpolating-hamiltonian}
\Hh^E(x)\;=\;\sum_{n=0}^N\Hh^E_n\,\chi(x\in(n-1,n])
\;,
\qquad
x\in[-1,N]
\;.
\end{equation}
The condition allows to extract a positive coefficient function $p$, and consecutively $q$ and $v$ from $\Hh^E$. These functions are piecewise continuous on $[-1,N]$. Hence, one can consider the associated fundamental solution $\Tt^E(x)$ obtained by solving \eqref{eq-HamSysGen}. On the interval $[n-1,n]$, the solution with initial condition $\one=\one_{2L}$ at $n-1$ is given by $x\in[n-1,n]\mapsto e^{(x-n+1)\Jj\Hh^E_n}$ so that at $x=n$ one has $e^{\Jj\Hh^E_n}=\Tt^E_n$. Hence starting with the left boundary condition $\Phi^E_{-1}=\Psi_D$, let us set
$$
U^E(x)
\;=\;
\Pi\big(\Tt^E(x)\Psi_D\big)
\;,
\qquad
x\in[-1,N]
\;.
$$
By the argument above, with this choice of $U^E(x)$, the Sturm oscillation \eqref{eq-SturmMatrixJac2} holds for $E< E_c$.

\vspace{.2cm}

It now remains to construct the selfadjoint $\Hh^E_n$ such that $\Tt_n^E=e^{\Jj\Hh_n^E}$ and the positivity \eqref{eq-HamSysPos} holds. Roughly stated, this means taking the logarithm of $\Tt_n^E$. As the transfer matrices are not normal and the logarithm has to satisfy the positivity condition, the functional calculus has to be carried out by hand, is somewhat lengthy and involves several steps.

\begin{proposi} 
\label{prop-TransferAnalysis} 
Let $V,T\in\CM^{L\times L}$ with $V=V^*$ and $T>0$.  Set 
\begin{equation}
\label{eq-TransferGen}
\Tt^E
\;=\;
\begin{pmatrix}
(E\,\one-V)T^{-1} & - T \\
T^{-1} & 0
\end{pmatrix}
\;\in\;\Gg(L)
\;.
\end{equation}
The spectrum of $\Tt^E$ lies in $\RM\cup\SM^1$ and is invariant under the map $\lambda\mapsto (\overline{\lambda})^{-1}$. Both eigenvalues $\lambda=-1$ and $\lambda=1$ always have even algebraic multiplicity with Jordan blocks of size $2$ with generalized eigenvectors. No other eigenvalue has a non-trivial Jordan block. As a function of $E$, all eigenvalue pairs $(\lambda^E,(\lambda^E)^{-1})$ move from the negative real axis via a Krein collision at $-1$ onto the unit circle and then they leave the unit circle again via a Krein collision at $1$.
\end{proposi}

\noindent {\bf Proof.} 
Let us begin by factorizing
$$
\Tt^E
\;=\;
\begin{pmatrix}
T^{-\frac{1}{2}} & 0 \\
0 & T^{\frac{1}{2}}
\end{pmatrix}^{-1}
\begin{pmatrix}
T^{-\frac{1}{2}}(E\,\one-V)T^{-\frac{1}{2}} & - \one \\
\one & 0
\end{pmatrix}
\begin{pmatrix}
T^{-\frac{1}{2}} & 0 \\
0 & T^{\frac{1}{2}}
\end{pmatrix}
\;.
$$
The next step is to diagonalize the selfadjoint matrix $T^{-\frac{1}{2}}(E\,\one-V)T^{-\frac{1}{2}}$ with a unitary matrix $M$, notably 
$$
M^*T^{-\frac{1}{2}}(E\,\one-V)T^{-\frac{1}{2}}M
\;=\;
D^E
\;,
$$ 
where $D^E$ is a real diagonal matrix. Of course, $M$ also depends on $E$, but this dependence is suppressed in the notations. As $T$ and thus also $T^{-1}$ are positive, this matrix $D^E$ is increasing in $E$. Now one has
\begin{equation}
\label{eq-UpperBlockDiag}
\begin{pmatrix}
T^{-\frac{1}{2}} & 0 \\
0 & T^{\frac{1}{2}}
\end{pmatrix}
\begin{pmatrix}
M & 0  \\ 0 & M
\end{pmatrix}
\Tt^E
\begin{pmatrix}
M & 0  \\ 0 & M
\end{pmatrix}^{-1}
\begin{pmatrix}
T^{-\frac{1}{2}} & 0 \\
0 & T^{\frac{1}{2}}
\end{pmatrix}^{-1}
\;=\;
\begin{pmatrix}
D^E & -\one  \\ \one & 0
\end{pmatrix}
\;.
\end{equation}
Hence, $\Tt^E$ is similar to a block diagonal real symplectic matrix with only $2\times 2$ blocks. Now all statements follow directly from the analysis of such $2\times 2$ blocks. While this is well-known, the main steps of this analysis are also contained in the proof of Proposition~\ref{prop-LogTransfer} below.
\hfill $\Box$

\vspace{.2cm}

Now it is possible to define the critical energy $E_c$ to be the smallest energy at which one of the transfer matrices $\Tt^E_n$, $n=1,\ldots,N$, undergoes a Krein collision at $1$. Alternatively,
$$
E_c
\;=\;
\sup\big\{E\in\RM\;:\;\sigma(\Tt^E_n)\subset (-\infty,0)\:\cup\:\SM^1\;\,\forall\;n=1,\ldots,N\big\}
\;.
$$

\vspace{.2cm}

The next result is now about functional calculus of $\Tt^E$. This is based on the diagonalization of $\Tt^E$ in the group $\Gg(L)=\{\Tt\in\CM^{2L\times 2L}:\Tt^*\Jj\Tt=\Jj\}$. On first sight, this merely looks like a corollary of the surjectivity of the exponential map for the Lie group $\Gg(L)$. However, the negativity claim on the lower right entry is a supplementary property that requires the use of the particular form of $\Tt^E$. 

\begin{proposi} 
\label{prop-LogTransfer} 
Let $\Tt^E$ be defined as in \eqref{eq-TransferGen} with $V=V^*$ and $T>0$.  Suppose that $E$ is such that the spectrum of $\Tt^E$ lies in $(-\infty,0)\cup (\SM^1\backslash\{1\})$. Then there exists $\Hh^E=(\Hh^E)^*\in\CM^{2L\times 2L}$  such that
$$
\Tt^E\;=\;e^{\Jj\Hh^E}
\;,
\qquad
\begin{pmatrix} 0 \\ \one \end{pmatrix}^*\Hh^E\begin{pmatrix} 0 \\ \one \end{pmatrix}
\;<\;0
\;.
$$
The map $E\mapsto \Hh^E$ is continuous.
\end{proposi}

\noindent {\bf Proof.} 
Let us start from the block diagonalization \eqref{eq-UpperBlockDiag} and also include a suitable permutation matrix in $M$ so that the eigenvalues of $D^E$ can be assumed to be increasing.  Next, note that the block-diagonal matrix $\Dd=\diag(T^{-\frac{1}{2}}M,T^{\frac{1}{2}}M)^{-1}$ is an element of the group $\Gg(L)$ and that the basis change with $\Dd$ does not alter the two required properties because
$$
\Dd^{-1}\,\Tt^E\,\Dd\;=\;e^{\Dd^{-1}\Jj\Hh^E\Dd}
\;=\;
e^{\Jj\Dd^*\Hh^E\Dd}
\;,
$$
and 
$$
\begin{pmatrix} 0 \\ \one \end{pmatrix}^*\Dd^*\,\Hh^E\,\Dd\begin{pmatrix} 0 \\ \one \end{pmatrix}
\;=\;
\big((T^{\frac{1}{2}}M)^{-1}\big)^*\begin{pmatrix} 0 \\ \one \end{pmatrix}^*\Hh^E\begin{pmatrix} 0 \\ \one \end{pmatrix}(T^{\frac{1}{2}}M)^{-1}
\;.
$$ 
Consequently, one can assume that $T^{\frac{1}{2}}M=\one$ or equivalently that $\Tt^E$ is given by the r.h.s. of \eqref{eq-UpperBlockDiag}. Next, let us introduce some notation by setting
$$
D^E
\;=\;
\diag(D^-_h,D_p^-,D_e,D_p^+,D_h^+)
\;,
$$ 
where the diagonal matrices $D^\pm_h$, $D_p^\pm$ and $D_e$ have sizes $L^\pm_h$, $L_p^\pm$ and $L_e$ respectively. Clearly
$$
L^-_h\,+\,L_p^-\,+\,L_e\,+\,L_p^+\,+\,L_h^+
\;=\;
L
\;.
$$
The indices $h$, $p$ and $e$ designate the hyperbolic, parabolic and elliptic blocks and the sizes are chosen such that 
$$
\pm \,D_h^\pm\;>\;2\,\one_{L_h^\pm}
\;,
\qquad
\pm \,D^\pm_p\;=\;2\,\one_{L_p^\pm}
\;,
\qquad
-2\,\one_{L_e}\;<\;D_e\;<\;2\,\one_{L_e}
\;.
$$
Let us note that all these sizes and diagonal matrices depend on $E$ in a controllable way. The matrices $D^\pm_h$ lead to hyperbolic $2\times 2$ blocks of $\Tt^E$ with real eigenvalues off the unit circle, while $D^\pm_p$ gives $2\times 2$ Jordan blocks of $\Tt^E$ with eigenvalues $\pm 1$ and finally $D_e$ leads to elliptic blocks which are similar to rotation matrices. The corresponding diagonalization procedures are now carried out in detail and in such a manner that all matrices are in the group $\Gg(L)$.  By the assumption of Proposition~\ref{prop-LogTransfer}, one has $L_p^+=L_h^+=0$. For further reference and because it is needed to explain the approach for large energies, we nevertheless first continue without this restriction.

\vspace{.1cm}

The next step is to perform the diagonalization procedures of the $2\times 2$ blocks in such a manner that all matrices are in the group $\Gg(L)$. Let us begin with the hyperbolic blocks. The corresponding eigenvalues of $\Tt^E$ are the diagonal entries of $\pm e^{\kappa^\pm}, \pm e^{-\kappa^\pm}$, where $\kappa^\pm>0$  of size $L_h^\pm$ is defined by
$$
e^{\pm\kappa^+}
\;=\;
\tfrac{D^+_h}{2}\,\pm\,
\big(\tfrac{(D_h^+)^2}{4}-\one\big)^{\frac{1}{2}}
\;,
\qquad
-\,e^{\pm\kappa^-}
\;=\;
\tfrac{D^-_h}{2}
\,\mp\,
\big(\tfrac{(D^-_h)^2}{4}-\one\big)^{\frac{1}{2}}
\;.
$$ 
The corresponding eigenvectors are given by
\begin{equation}
\label{eq-D-h}
\begin{pmatrix}
D_h^\pm & -\one  \\ \one & 0
\end{pmatrix}
\begin{pmatrix}
\pm e^{\kappa^\pm} & e^{-\kappa^\pm} \\ \one & \pm\one
\end{pmatrix}
\;=\;
\begin{pmatrix}
\pm e^{\kappa^\pm} & e^{-\kappa^\pm} \\ \one & \pm\one
\end{pmatrix}
\begin{pmatrix}
\pm e^{\kappa^\pm} & 0 \\ 0 & \pm e^{-\kappa^\pm} 
\end{pmatrix}
\;.
\end{equation}
One can also choose the matrix of eigenvectors to be in the group $\Gg(L^\pm_h)$ associated to $\Jj_h^\pm$ (namely the subset of $\CM^{2L^\pm_h\times 2L^\pm_h}$ which conserve $\Jj_h^\pm$ as a quadratic form). This is achieved by normalizing each matrix entry with the inverse square root of $e^{\kappa^\pm}-e^{-\kappa^\pm}>0$, which leads to set
$$
(\Mm^\pm_h)^{-1}
\;=\;
\begin{pmatrix}
\frac{\pm e^{\kappa^\pm}}{(e^{\kappa^\pm}-e^{-\kappa^\pm})^{\frac{1}{2}}} & \frac{e^{-\kappa^\pm}}{ (e^{\kappa^\pm}-e^{-\kappa^\pm})^{\frac{1}{2}}}\\ 
\frac{1}{(e^{\kappa^\pm}-e^{-\kappa^\pm})^{\frac{1}{2}}} & \frac{\pm 1}{(e^{\kappa^\pm}-e^{-\kappa^\pm})^{\frac{1}{2}}}
\end{pmatrix}
\;\in\;\Gg(L^\pm_h)
\;.
$$
Let us stress that these matrices diverge as $\kappa^\pm\to 0$, namely one approaches a Jordan block, but this divergence will disappear once the Hamiltonian is computed.

\vspace{.1cm}

As to the elliptic block, the eigenvalues on the upper half of the unit circle are given by 
$$
-e^{\imath\theta}
\;=\;
\tfrac{D_e}{2}\,+\imath\,\big(\one-\tfrac{D_e^2}{4}\big)^{\frac{1}{2}}
\;,
\qquad
-e^{-\imath\theta}
\;=\;
\tfrac{D_e}{2}\,-\,
\imath\big(\one-\tfrac{D_e^2}{4}\big)^{\frac{1}{2}}
\;.
$$  
Here, $\theta$ is chosen to have diagonal entries in $(-\pi,0)$. The $2\times 2$ block corresponding to $D_e$ can be diagonalized exactly as in \eqref{eq-D-h}, but neither the resulting basis change nor the diagonal matrix with complex entries are in the group $\Gg(L_e)$ of matrices in $\CM^{2L_e\times 2L_e}$ conserving $\Jj_e$. To achieve the latter, one rather transforms into a rotation matrix:
$$
\begin{pmatrix}
D_e & -\one  \\ \one & 0
\end{pmatrix}
\begin{pmatrix} \cos\theta(-\sin\theta)^{-\frac12} & (-\sin\theta)^\frac12 \\ -(-\sin\theta)^{-\frac12} & 0\end{pmatrix}
\;=\;
\begin{pmatrix} \cos\theta & -\sin\theta \\ -\one & 0\end{pmatrix}\frac1{\sqrt{-\sin\theta}}
\begin{pmatrix}
-\cos \theta & \sin\theta
\\
-\sin\theta & -\cos\theta 
\end{pmatrix}
\;.
$$
Hence, let us set
$$
(\Mm_e)^{-1}
\;=\;
\begin{pmatrix}
\cos\theta(-\sin\theta)^{-\frac12} & (-\sin\theta)^\frac12 \\ 
-(-\sin\theta)^{-\frac12} & 0
\end{pmatrix}\;\in\;\Gg(L_e)
\;.
$$
Finally, the parabolic cases are based on the identities
\begin{equation}
\label{eq:parabolic-log-neg}
\begin{pmatrix}-2\,\one & -\one \\ \one & 0 \end{pmatrix}
\;=\;
-\left(\begin{pmatrix}\one & 0 \\ 0 & \one \end{pmatrix} + \begin{pmatrix} \one & \one \\ -\one & -\one \end{pmatrix}\right)
\;=\;
\exp\left(\begin{pmatrix}(1+\imath\pi)\one & \one \\ -\one & (-1+\imath\pi)\one \end{pmatrix}\right)
\;,
\end{equation}
and
\begin{equation}
\label{eq:parabolic-log-pos}
\begin{pmatrix} 2\,\one & -\one \\ \one & 0 \end{pmatrix}
\;=\;
\begin{pmatrix}\one & 0 \\ 0 & \one \end{pmatrix} + \begin{pmatrix} \one & -\one \\ \one & -\one \end{pmatrix}
\;=\;
\exp\left(\begin{pmatrix}\one & -\one \\ \one & -\one \end{pmatrix}\right)
\;.
\end{equation}
Hence let us also set $(\Mm_p^\pm)^{-1}=\one$. To regroup all the above, it is convenient to use the notation of diagonal checkerboard sums:
$$
\begin{pmatrix} A_1 & B_1 \\ C_1 & D_1 \end{pmatrix}\widehat{\oplus}\begin{pmatrix} A_2 & B_2 \\ C_2 & D_2 \end{pmatrix}
\;=\;
\begin{pmatrix}
A_1 & 0 & B_1 & 0 \\
0 & A_2 & 0 & B_2 \\
C_1 & 0 & D_1 & 0 \\
0 & C_2 & 0 & D_2 \\
\end{pmatrix}
\;.
$$
Then, the basis changes can be collected as
$$
\Mm^{-1}
\;=\;
(\Mm_h^-)^{-1}\:\widehat{\oplus}\:(\Mm_p^-)^{-1}\:\widehat{\oplus}\:(\Mm_e)^{-1}\:\widehat{\oplus}\:(\Mm_p^+)^{-1}\:\widehat{\oplus}\:(\Mm_h^+)^{-1}
\;,
$$
and one also has
$$
\Jj
\;=\;
\Jj_h^-\:\widehat{\oplus}\:\Jj_p^-\:\widehat{\oplus}\:\Jj_{e}\:\widehat{\oplus}\:\Jj_p^+\:\widehat{\oplus}\:\Jj_h^+
\;.
$$
Note that $\Gg(L^-_{h})\:\widehat{\oplus}\:\Gg(L_p^-)\:\widehat{\oplus}\:\Gg(L_{e})\:\widehat{\oplus}\:\Gg(L_p^+)\:\widehat{\oplus}\:\Gg(L^+_h)$ is a subgroup of $\Gg(L)$, which is strict except in the trivial case. Then $\Mm^{-1}\in\Gg(L)$ and thus $\Mm=\Jj^*(\Mm^{-1})^*\Jj$ is given by
$$
\Mm
\;=\;
\Mm^-_h\:\hat{\oplus}\Mm^-_p\:\hat{\oplus}\:\Mm_e\:\hat{\oplus}\:\Mm_p^+\:\hat{\oplus}\:\Mm^+_h
\;,
$$
with summands 
$$
\Mm^\pm_h
\;=\;
\begin{pmatrix}
\frac{\pm 1}{ (e^{\kappa^\pm}-e^{-\kappa^\pm})^{\frac{1}{2}}} & 
\frac{-e^{-\kappa^\pm}}{ (e^{\kappa^\pm}-e^{-\kappa^\pm})^{\frac{1}{2}}}
\\ 
\frac{-1}{(e^{\kappa^\pm}-e^{-\kappa^\pm})^{\frac{1}{2}}} 
& \frac{\pm e^{\kappa^\pm}}{(e^{\kappa^\pm}-e^{-\kappa^\pm})^{\frac{1}{2}}}
\end{pmatrix}
\;,
\qquad
\Mm_e
\;=\;
\begin{pmatrix}
0 & -(-\sin\theta)^\frac12 \\
(-\sin\theta)^{-\frac{1}{2}} & \cos\theta(-\sin\theta)^{-\frac12}
\end{pmatrix}\;,
$$
as well as $\Mm_p^\pm=\one$. Furthermore $\Nn=\Mm{\Tt}^E\Mm^{-1}$ is given by
$$
\begin{pmatrix}
-e^{\kappa^-} & 0 \\ 0 & -e^{-\kappa^-} 
\end{pmatrix}
\,\hat{\oplus}\,
\begin{pmatrix}
-2\one & -\one \\
\one & 0
\end{pmatrix}
\,\hat{\oplus}\,
\begin{pmatrix}
-\cos\theta & \sin\theta
\\
-\sin\theta & -\cos\theta 
\end{pmatrix}
\,\hat{\oplus}\,\begin{pmatrix}
2\one & -\one \\
\one & 0
\end{pmatrix}
\,\hat{\oplus}\,
\begin{pmatrix}
e^{\kappa^+} & 0 \\
 0 & e^{-\kappa^+} 
\end{pmatrix}
\;.
$$
One can now (using equations \eqref{eq:parabolic-log-neg} and \eqref{eq:parabolic-log-pos} for the parabolic cases) readily take the (principal branch of the) logarithm $\mbox{\rm Log}(\Nn)$ such that $\Nn=\exp(\mbox{\rm Log}(\Nn))$, namely
\begin{align*}
\mbox{\rm Log}(\Nn)
\;=\;&
\begin{pmatrix}
\kappa^-  +\imath \pi\one & 0 
\\ 
0 & -\kappa^- +\imath \pi \one
\end{pmatrix}
\,\widehat{\oplus}\,
\begin{pmatrix}
(1+\imath\pi)\one & \one \\
-\one & (-1+\imath\pi)\one
\end{pmatrix}
\,\widehat{\oplus}\,
\begin{pmatrix}
\imath\pi\one  & - \theta 
\\ 
\theta & \imath\pi\one
\end{pmatrix}
\\
&
\;\;
\widehat{\oplus}\,
\begin{pmatrix}
\one & -\one \\
\one & -\one
\end{pmatrix}
\,\widehat{\oplus}\,
\begin{pmatrix}
\kappa^+ & 0  \\ 
0 & -\kappa^+
\end{pmatrix}
\;.
\end{align*}
Now $\Jj\Jj^*=\one$ implies
$$
{\Tt}^E
\;=\;
\Mm^{-1}\,\exp(\mbox{\rm Log}(\Nn))\,\Mm
\;=\;
\exp
\big(
\Jj\Mm^*\Jj^*\mbox{\rm Log}(\Nn)\Mm\big)
\;.
$$
Hence, the Hamiltonian is ${\Hh}^E=\Mm^*\Jj^*\mbox{\rm Log}(\Nn)\Mm$ and given by
\begin{align*}
{\Hh}^E
\;=\;
\Mm^*
\;
&
\begin{pmatrix}
0 & -\kappa^-  +\imath \pi\one
\\ 
-\kappa^-  -\imath \pi\one & 0
\end{pmatrix}
\,\hat{\oplus}\,
\begin{pmatrix}
-\one & (-1+\imath\pi)\one \\
(-1-\imath\pi)\one & -\one
\end{pmatrix}
\,\hat{\oplus}\,
\begin{pmatrix}
\theta & \imath\pi\one 
\\ 
-\imath\pi\one & \theta 
\end{pmatrix}\,
\\
&
\,\hat{\oplus}\,
\begin{pmatrix}
\one & -\one \\
-\one & \one
\end{pmatrix}
\,\hat{\oplus}\,
\begin{pmatrix}
0 & -\kappa^+
\\ 
-\kappa^+ & 0
\end{pmatrix}
\;
\Mm
\;.
\end{align*}
As $\Mm$ is also checkerboard diagonal, one can now check that 
$$
{\Hh}^E
\;=\;
{\Hh}^{E,-}_h
\,\hat{\oplus}\,
{\Hh}^{E,-}_{p}
\,\hat{\oplus}\,
{\Hh}^{E}_{e}
\,\hat{\oplus}\,
{\Hh}^{E,+}_{p}
\,\hat{\oplus}\,
{\Hh}^{E,+}_{h}
$$
with
\begin{align*}
{\Hh}^{E,-}_{h}
& 
\;=\;
\begin{pmatrix} \frac{-\kappa^-}{\sinh\kappa^-} & \frac{-\kappa^-\cosh\kappa^-}{\sinh\kappa^-}+\imath\pi\one \\
\frac{-\kappa^-\cosh\kappa^-}{\sinh\kappa^-}-\imath\pi\one & \frac{-\kappa^-}{\sinh\kappa^-}
\end{pmatrix}\;,
\\
{\Hh}^{E,-}_p
&  \;=\;
\begin{pmatrix}
-\one & (-1+\imath\pi)\one \\
(-1-\imath\pi)\one& -\one
\end{pmatrix}
\;,
\\
{\Hh}^{E}_{e}
\;\;\,
& \;=\;
\begin{pmatrix}
-\frac{\theta}{\sin\theta} & 
\frac{-\theta\cos\theta}{\sin\theta}+\imath\pi\one
\\ 
\frac{-\theta\cos\theta}{\sin\theta}-\imath\pi\one & 
-\frac{\theta}{\sin\theta}
\end{pmatrix}
\;,
\addtocounter{equation}{1}\tag{\theequation}
\label{eq-hamiltonian-formulas}
\\
\Hh^{E,+}_p
& \;=\;
\begin{pmatrix}
\one & -\one \\
-\one & \one
\end{pmatrix},\\
\Hh^{E,+}_h
& 
\;=\;
\begin{pmatrix}
 \frac{\kappa^+}{\sinh\kappa^+} &  
\frac{-\kappa^+\cosh\kappa^+}{\sinh\kappa^+}
\\ 
\frac{-\kappa^+\cosh\kappa^+}{\sinh\kappa^+} & 
\frac{\kappa^+}{\sinh\kappa^+} \end{pmatrix} 
\;.
\end{align*}
Now due to the assumptions on the spectrum of $\Tt^E$ in Proposition~\ref{prop-LogTransfer}, there are only the first three summands for which one now simply reads off the desired negativity property. Regarding continuity, $\kappa^\pm$ and $\theta$ depend continuously on the diagonal entries of $D^E$, which in turn depend continuously on $E$ by \eqref{eq-UpperBlockDiag}, so continuity is preserved within each block. Finally, one observes that as an eigenvalue moves toward the Krein collision at $-1$ from the real line (resp. the circle), the limit of the corresponding blockdiagonal entries of the negative hyperbolic (resp. elliptic) Hamiltonian block converge precisely toward the corresponding entries of $\Hh_p^{E,-}$.
\hfill$\Box$.

\vspace{.2cm}

\noindent {\bf Remark} Considering the behavior as the eigenvalues of $\Tt^E$ move toward the Krein collision at $1$, the entries of $\theta$ tend toward $-\pi$ and thus the entries of $\Hh^E_e$ in \eqref{eq-hamiltonian-formulas} diverge. Due to the properties of the logarithm, some form of either divergence or discontinuity of $\Hh^E$ is unavoidable as soon as the spectrum of $\Tt^E$ is allowed to include the entire circle. This is why $E$ being less than the critical energy $E_c$ has to be imposed for the homotopy argument performed above. However, one can choose the divergence to be at the Krein collision at $-1$ while preserving continuity at the Krein collision at $1$ as follows:  With $D_e$ as above, choose $\theta$ to have diagonal entries in $(0,\pi)$ such that
$$
e^{\imath\theta}
\;=\;
\tfrac{D_e}{2}\,+\,\imath\big(\one-\tfrac{(D_e)^2}{4}\big)^{\frac{1}{2}}
\;,
\qquad
e^{-\imath\theta}
\;=\;
\tfrac{D_e}{2}\,-\,
\imath{\big(\one-\tfrac{(D_e)^2}{4}\big)}^{\frac{1}{2}}
\;.
$$  
Then one now obtains
$$
\begin{pmatrix}
D_e & -\one  \\ \one & 0
\end{pmatrix}
\begin{pmatrix}
(\sin\theta)^{\frac{1}{2}} & \cos\theta(\sin\theta)^{-\frac{1}{2}}
\\
0 & (\sin\theta)^{-\frac{1}{2}} 
\end{pmatrix}
\;=\;
\begin{pmatrix}
(\sin\theta)^{\frac{1}{2}} &  \cos\theta(\sin\theta)^{-\frac{1}{2}}
\\
0 & (\sin\theta)^{-\frac{1}{2}} 
\end{pmatrix}
\begin{pmatrix}
\cos\theta & -\sin \theta\\
\sin\theta & \cos\theta
\end{pmatrix}
\;
$$
and
$$
(\Mm_e)^{-1}\;=\;\begin{pmatrix}
(\sin\theta)^\frac12 & \cos\theta(\sin\theta)^{-\frac12} \\
0 & (\sin\theta)^{-\frac12}
\end{pmatrix}
\;\in\;\Gg(L_e)
\;.
$$
As above, one can thus calculate the Hamiltonian for the elliptic block as
$$
\Hh^{E}_e
\;=\;
(\Mm_e)^*\Jj^*\begin{pmatrix} 0 & -\theta \\ \theta & 0 \end{pmatrix}\Mm_e
\;=\;
\begin{pmatrix} \frac{\theta}{\sin\theta} & -\frac{\theta\cos\theta}{\sin\theta} \\
-\frac{\theta\cos\theta}{\sin\theta} & \frac{\theta}{\sin\theta}
\end{pmatrix}
\;.
$$
Note that now the lower right entry of $\Hh^E_e$ is positive, just as for $\Hh_p^{E,+}$ and $\Hh_h^{E,+}$ as calculated in \eqref{eq-hamiltonian-formulas}. Furthermore, with this choice of the Hamiltonian of the elliptic block, continuity in $E$ is preserved in the positive Krein collision, while the divergence occurs at the Krein collision at $-1$. Having the same signs, one can now construct $U^E(x)$ and for energies larger than 
$$
E_c^\prime
\;=\;
\inf\big\{E\in\RM\;:\;\sigma(\Tt^E_n)\subset (0,\infty)\:\cup\:\SM^1\;\,\forall\;n=1,\ldots,N\big\}
\;,
$$
one can adapt (with some effort) the homotopy argument so that \eqref{eq-SturmMatrixJac2} holds.
\hfill $\diamond$

\vspace{.3cm}

\noindent {\bf Acknowledgements:} We thank Roman Hilscher for explaining discrete Sturm oscillations to us which ultimately lead to Section~\ref{sec-PrueferJacobi}. This work was partially supported by the DFG. 



\end{document}